\documentclass[10pt,journal,compsoc]{IEEEtran}
\ifCLASSOPTIONcompsoc
  \usepackage[nocompress]{cite}
\else
  \usepackage{cite}
\fi
\ifCLASSINFOpdf
  \usepackage[pdftex]{graphicx}
\else
  \usepackage[dvips]{graphicx}
\fi

\usepackage{algorithmic}
\usepackage{array}

\ifCLASSOPTIONcompsoc
 \usepackage[caption=false,font=footnotesize,labelfont=sf,textfont=sf]{subfig}
\else
 \usepackage[caption=false,font=footnotesize]{subfig}
\fi
\usepackage{stfloats}

\ifCLASSOPTIONcaptionsoff
 \usepackage[nomarkers]{endfloat}
\let\MYoriglatexcaption\caption
\renewcommand{\caption}[2][\relax]{\MYoriglatexcaption[#2]{#2}}
\fi

\let\MYorigsubfloat\subfloat
\renewcommand{\subfloat}[2][\relax]{\MYorigsubfloat[]{#2}}

\usepackage{times}
\usepackage{epsfig}
\usepackage{graphicx}
\usepackage{amsmath,bm}
\usepackage{amssymb}
\usepackage{booktabs}
\usepackage{multirow}
\usepackage{color}
\usepackage{array}
\usepackage{multirow}
\usepackage{multicol}
\usepackage[dvipsnames]{xcolor}
\usepackage{algorithmic}
\usepackage{amsmath,tikz}
\usepackage[english]{babel}
\usepackage{amsfonts}
\usepackage[ruled]{algorithm2e} %
\usepackage{stfloats}

\usepackage[caption=false,font=footnotesize,labelfont=sf,textfont=sf]{subfig}
\usepackage[nocompress]{cite}

\let\MYorigsubfloat\subfloat
\renewcommand{\subfloat}[2][\relax]{\MYorigsubfloat[]{#2}}

\newcommand{\yjtvcg}[1]{{\color{black}#1}}
\newcommand{\yj}[1]{{\color{black}#1}}
\newcommand{\yjrev}[1]{{\color{black}#1}}
\newcommand{\gl}[1]{{\color{black}#1}}
\newcommand{\YL}[1]{{\color{black}#1}}
\newcommand{\error}[1]{{\color{black}#1}}
\newcommand{\hyh}[1]{{\color{black}#1}}

\DeclareMathOperator*{\argmax}{arg\!\max}

\newcommand{\PreserveBackslash}[1]{\let\temp=\\#1\let\\=\temp}
\newcommand{\tabincell}[2]{\begin{tabular}
{@{}#1@{}}#2\end{tabular}}
\newcolumntype{C}[1]{>{\PreserveBackslash\centering}p{#1}}
\newcolumntype{R}[1]{>{\PreserveBackslash\raggedleft}p{#1}}
\newcolumntype{L}[1]{>{\PreserveBackslash\raggedright}p{#1}}
\makeatletter
\newcommand\figcaption{\def\@captype{figure}\caption}
\newcommand\tabcaption{\def\@captype{table}\caption}
\makeatother

\begin{document}

\title{Multiscale Mesh Deformation Component Analysis with Attention-based Autoencoders}

\author{Jie~Yang,%
        ~Lin~Gao\thanks{\IEEEauthorrefmark{1} Corresponding author is Lin Gao.}\IEEEauthorrefmark{1},%
        ~Qingyang~Tan,%
        ~Yihua~Huang,
        ~Shihong~Xia
        ~and~Yu-Kun~Lai
\IEEEcompsocitemizethanks{
\IEEEcompsocthanksitem J. Yang, L. Gao, Y. Huang and S. Xia are with the Beijing Key Laboratory of Mobile Computing and Pervasive Device, Institute of Computing Technology, Chinese Academy of Sciences and also with the University of Chinese Academy of Sciences, Beijing, China. \protect\\
E-mail: \{yangjie01, gaolin, huangyihua20g, xsh\}@ict.ac.cn \protect\
\IEEEcompsocthanksitem Q. Tan is with University of Maryland, College Park, Maryland, USA.\protect\\
E-mail: qytan@cs.umd.edu \protect\
\IEEEcompsocthanksitem Y.-K Lai is with Visual Computing Group, School of Computer Science and Informatics, Cardiff University, Wales, UK.\protect\\
E-mail: LaiY4@cardiff.ac.uk
}%
\thanks{Manuscript received April 19, 2019; revised August 26, 2019.}}

\markboth{IEEE TRANSACTIONS ON VISUALIZATION AND COMPUTER GRAPHICS,~Vol.~xx, No.~xx, July~2019}%
{Yang \MakeLowercase{\textit{et al.}}: Multiscale Mesh Deformation Component Analysis with Attention based Autoencoders}

\IEEEtitleabstractindextext{%
\begin{abstract}
  \yj{Deformation component analysis is a fundamental problem in geometry processing and shape understanding. Existing approaches mainly extract deformation components in local regions at a similar scale while deformations of real-world objects are usually distributed in a multi-scale manner. In this paper, we propose a novel method to exact multiscale deformation components automatically with a stacked attention-based autoencoder. The attention mechanism is designed to learn to softly weight multi-scale deformation components in active deformation regions, and the stacked attention-based autoencoder is learned to represent the deformation components at different scales. Quantitative and qualitative evaluations show that our method outperforms state-of-the-art methods.
  Furthermore, with the multiscale deformation components extracted by our method, the user can edit shapes in a coarse-to-fine fashion which facilitates effective modeling of new shapes.}
\end{abstract}

}

\maketitle

\IEEEdisplaynontitleabstractindextext

\IEEEpeerreviewmaketitle

\ifCLASSOPTIONcompsoc
\IEEEraisesectionheading{\section{Introduction}\label{sec:introduction}}
\else
\section{Introduction}
\label{sec:introduction}
\fi
\IEEEPARstart{W}{ith} the development of 3D scanning and modeling technology, %
3D mesh collections are becoming much more popular. These mesh models usually use fixed vertex connectivity with variable vertex positions to characterize different shapes. 
Analyzing these mesh model collections to extract meaningful components and using these components for new model generation are key research problems in these areas.   
Some works~\cite{neumann2013sparse,huang2014sparse,wang2017articulated,sparsevae2017} propose to extract deformation components from mesh data sets. They mainly focus on extracting local deformation components with sparse regularization \gl{at a uniform scale}. However, real-world objects deform at multiple scales. For example, a person may have different facial expressions which are more localized deformations on the face, %
\gl{but the whole body can also be bent, which is a larger scale deformation.}

\yjrev{Multi-scale techniques are getting increasingly popular in various fields. In Finite Element Methods, multiscale analysis is widely used~\cite{alleman2018concurrent,mathew2018multiscale,abeyratne2002multiscale}. In the spectral geometry field, research works apply multiscale technology on the deformation representation~\cite{lam2017multiscale}, physics-based simulation of deformable objects~\cite{yang2013boundary}, and surface registration~\cite{hamidian2019surface} by analyzing the non-isometric global and local (multiscale) deformation. \hyh{Moreover, for shape editing, multiscale technology also enables modeling rich facial expressions on human faces~\cite{liu2020ras}.}}

Such multiscale deformation components are especially useful to support model editing from coarse level to fine level.
\yj{One motivation of this work is to achieve %
editing consistent with perceptual semantics by modifying shapes at suitable scales. The user would be able to make rough editing of the overall shape at a large scale, as well as localized modifications to surface details at a small scale.} %
Inspired by the recent advances in image processing with attention mechanism~\cite{fu2017look}, the attention is formulated to focus on specific regions in our approach.

\begin{figure*}[!th]
\begin{center}
\includegraphics[width=0.99\linewidth]{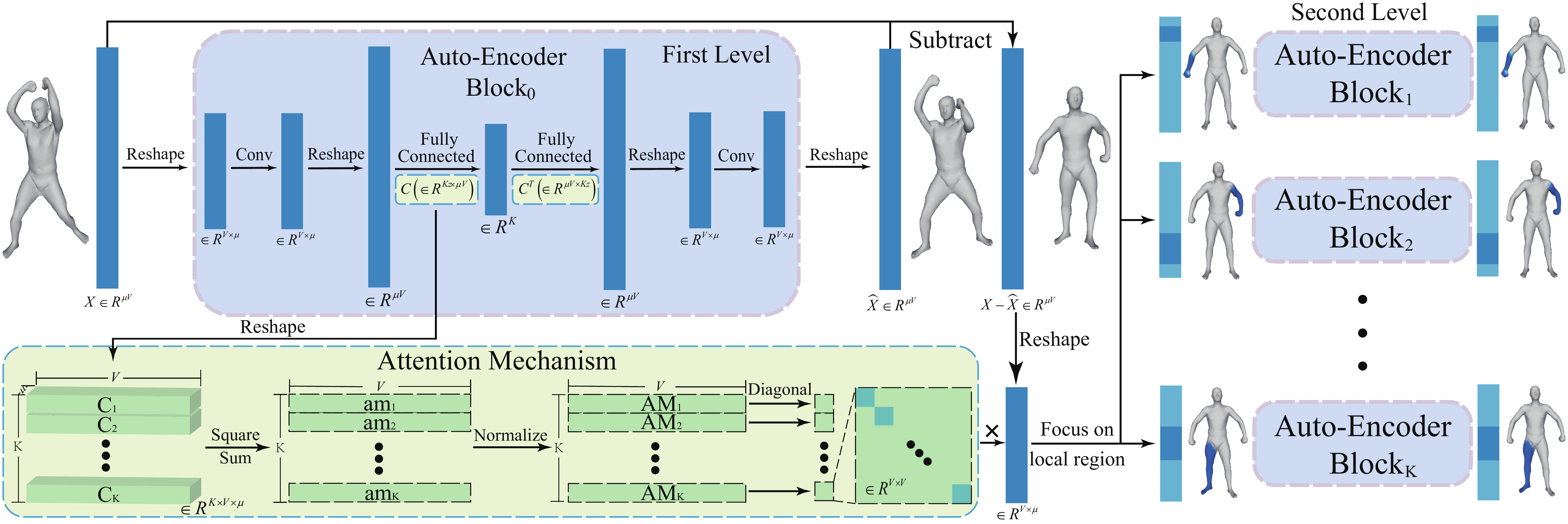}
\end{center}%
\caption{Our network architecture with attention mechanism and stacked autoencoders. We obtain large-scale deformation components and attention masks from the first-level autoencoder $AE_0$. For the second-level autoencoders $AE_k, 1\le k\le K$, we put the residual value $X-\widehat{X}$ of the first-level autoencoder $AE_0$ with the attention masks into $K$ autoencoders focusing on different sub-regions of the shape. We can then obtain small-scale deformation components by these autoencoders. This architecture can be further extended to include more scale levels.
The autoencoder has mirrored encoder and decoder structure. The encoder has one convolution layer and a fully connected layer, and the encoder and decoder share the same trainable parameters. 
$K$, $V$ and $\mu$ are the dimension of the latent space (the number of attention masks),
the number of  vertices and the dimension of the vertex features, respectively. By making the latent vector as a one-hot vector, we can extract the $K$ attention masks $AM_{k},1\le k\le K$ from the parameter $C$ as the top-left corner of the figure shows. Please refer to Sec.~\ref{sec:attention} for details. 
}
\label{fig:network}
\end{figure*}

We propose a novel autoencoder architecture to extract  multiscale local deformation components from shape deformation datasets. Our network structure is based on the mesh-based convolutional autoencoder architecture and also uses an effective representation of the shapes~\cite{gao2017sparse} as input which is able to encode large-scale deformations. 
\hyh{In this work, a stacked autoencoder architecture is proposed such that the network can encode the residual value of the former autoencoder with the attention mechanism, which helps to seperate the deformations into different scales and extract multiscale local deformation components.} %
The network architecture is shown in Fig.~\ref{fig:network}. We further utilize a sparsity constraint on the parameters of the fully connected layers to keep the deformation components localized. The autoencoder architecture ensures the extracted deformation components are suitable for multiscale shape editing and helps reconstruct high quality shapes  with less distortion. 

Our contributions are twofold:
\begin{itemize}
\item To the best of our knowledge, 
this is the first work that automatically extracts multiscale deformation components from a deformed shape collection. With these extracted components, the user can edit the 3D mesh shape in a coarse-to-fine fashion, which makes 3D modeling much more effective.

\item \yjtvcg{To achieve this, we propose a novel deep architecture involving attentional stacked autoencoders.
The attention mechanism is designed for learning the soft weights that help extract multiscale deformation components in the shape analysis and the stacked autoencoders are used to decompose the deformation of shape collections into different shape components with different scales.}
\end{itemize}

All the components of the network are tightly integrated and help each other. The attention mechanism makes the follow-up autoencoders focus on a specific region, to allow extracting smaller-scale local deformation components. Extensive comparisons prove that our method extracts more meaningful deformation components than state-of-the-art methods.

In Sec.~\ref{sec:related}, we review the most related work. We then give a brief description of the input features used in our method in Sec.~\ref{sec:feature}, and present
detailed description of our novel autoencoder with the attention mechanism including implementation details in Sec.~\ref{sec:network}. %
\yjtvcg{Finally, we present experimental results, including extensive comparisons with state-of-the-art methods in Sec.~\ref{sec:result} \yjtvcg{and draw conclusions in Sec.~\ref{sec:conclusion}}. }

\section{Related Work}\label{sec:related}

\yjtvcg{Mesh deformation component analysis has attracted significant interest in the research of shape analysis and data-driven mesh deformation. Many data-driven methods for editing of either man-made objects~\cite{xu2009joint,yumer2014co,yumer2015semantic} or general deformable surfaces~\cite{zhou2005large,gao2016efficient,gao2017sparse} benefit from extracted deformation components.
\hyh{Our work shares the same interest as theirs, aiming to assist users to edit shapes efficiently. Although our focus is to extract more meaningful multiscale deformation components automatically, it can be incorporated into existing data-driven deformation methods.} %
In the following, we will review work most related to ours. } 

\textbf{3D shape deformation component extraction.}
With the increasing use of 3D models, the need for analyzing their intrinsic characteristics becomes mainstream. Early work~\cite{alexa2000representing} extracts principal components from the mesh data set by Principal Component Analysis (PCA), but the extracted components contain %
\gl{global deformations}, which is not effective for users to make local edits. Some works~\cite{Gao2012} demonstrate that sparse constraints are effective for achieving localized deformation results. However, the classical sparse PCA~\cite{Zou2004} does not take the spatial information into consideration. By promoting sparsity in the spatial domain, many works extract localized deformation components with a sparsity constraint~\cite{neumann2013sparse, bernard2016linear}, which outperforms the standard PCA variants such as Clustering-PCA~\cite{tena2011interactive} with respect to choosing suitable compact basis modes, especially for producing more localized meaningful deformation. Moreover, the pioneering work~\cite{neumann2013sparse} represents meshes with Euclidean coordinates, but this representation is sensitive to rigid and non-rigid transformations. Later work~\cite{huang2014sparse} extends the previous work~\cite{neumann2013sparse} to better deal with  rotations by using deformation gradients to represent shapes, but the method still cannot cope with larger rotations greater than $180^{\circ}$ due to their inherent ambiguity.  
\yjrev{Based on deformation gradients, Neumann et al.~\cite{neumann2013capture} learn the arm-muscle deformation using a small set of intuitive parameters.}
The work~\cite{wang2017articulated} extends~\cite{neumann2013sparse} by using a rotation invariant representation based on edge lengths and dihedral angles~\cite{frohlich2011example}, so can handle large-scale deformations. 
\yjtvcg{However, the representation~\cite{frohlich2011example} is not suitable for extrapolation as this would result in negative edge lengths. 
This limits the capability of~\cite{wang2017articulated} for deformation component analysis, \YL{as extrapolation is often needed e.g. when utilizing the extracted deformation components for data-driven shape editing.}} 
Recent work~\cite{sparsevae2017} proposes a convolutional autoencoder based on an effective shape representation~\cite{gao2017sparse} to learn the localized deformations of a shape set, but their architecture is not suitable for extracting multiscale deformation components. 
Overall, different from these works~\cite{neumann2013sparse, sparsevae2017, huang2014sparse, wang2017articulated}, our method can produce meaningful and multiscale localized deformation components. %

\textbf{Deep learning on 3D Shapes.}
With the development of artificial intelligence, deep learning and neural networks have made great progress in many areas, in particular 2D image processing. 
\yjrev{Hence, some researchers transform the non-uniform geometry signals defined on meshes of different topologies into a regular domain, while preserving shape information as much as possible, which enables powerful uniform Cartesian grid based CNN (Convolutional Neural Network) backbone architectures to be used on problems such as cross-domain shape retrieval~\cite{chen2020cross}, surface reconstruction~\cite{jiang2019convolutional} and shape completion~\cite{sarkar20183d}. DDSL~\cite{jiang2019ddsl} was recently proposed, which is a differentiable layer compatible with deep neural networks for learning geometric signals.}
However, due to the irregularity of 3D shapes, deep learning is difficult to apply straightaway. Inspired by image processing, \yjtvcg{some works~\cite{maturana2015voxnet,su2015multi,yumer2016learning}} apply deep learning to the voxel representation with regular connectivity. However, voxel representation incurs significant computation and memory costs, which limits the resolution such methods can handle.
Wang et al.~\cite{wang2017cnn,wang2018adaptive} improve the performance of voxel-based convolutions by proposing an adaptive octree structure to represent 3D shapes, and apply it to shape completion~\cite{wang2020deep}. Meanwhile, recent works~\cite{li2018pointcnn, lei2018spherical} define the convolution on point clouds by using K-nearest-neighbors (KNN) and spherical convolutions. \yjrev{More recently, the work~\cite{wang2019dynamic} proposed the EdgeConv operator for learning on the point cloud to improve the performance of segmentation and classification.} In addition to the voxel and point cloud representations, shapes can also be represented as multiview projection images to perform 2D-CNNs~\cite{shi2015deeppano,su2015multi,sarkar2018learning} for 3D object recognition and classification. Such approaches are used to learn the local shape descriptors for shape segmentation and correspondence~\cite{huang2018learning}. \YL{For applications that take meshes as input or generate meshes as output, turning meshes to alternative representations can lead to useful topology information to be lost.}

Alternatively, as a mesh can be represented as a graph, CNNs can be extended to graph CNNs in the spatial domain~\cite{duvenaud2015convolutional, niepert2016learning} or spectral domain~\cite{henaff2015deep, bruna2013spectral, defferrard2016convolutional} for mesh-based deep learning. 
\yjrev{In the spatial domain, works~\cite{meshvae2017, COMA:ECCV18, Tretschk2019arXiv, litany2018deformable} apply variational autoencoders on 3D meshes for various applications such as reconstruction, interpolation, completion and embedding. The work~\cite{fulton2019latent} uses autoencoders to analyze  deformable solid dynamics.} However, none of the existing methods can extract multiscale deformation components, which we address by using a novel attention-based mesh convolutional network architecture.

\textbf{Attention mechanism on convolutional networks.}
Deep neural networks have proved their superiority in extracting high-level semantics and highly discriminative features on various image datasets. Researchers now pay more attention to using convolution features more effectively on fine-grained datasets to improve the performance. Such works can be widely seen in different areas of computer vision and natural language processing, such as image translation (DA-GAN)~\cite{ma2018gan}, person re-identification~\cite{si2018dual,xu2018attention}, document classification~\cite{yang2016hierarchical}, object detection~\cite{zhang2018progressive,zhuang2018parallel}, video classification~\cite{long2018attention}, etc. The work~\cite{bahdanau2014neural} proposes a ``soft attention'' mechanism which predicts soft weights and computes a weighted combination of the items in machine translation. In ~\cite{lu2016hierarchical}, a hierarchical co-attention method is proposed to learn the conditional representation of the image given a problem. Following~\cite{lu2016hierarchical}, \cite{wang2017vqa} extends the co-attention model to higher orders. Some works effectively utilize attention as a way to focus on specific regions for learning. Wang et al.~\cite{wang2017residual} demonstrate the benefit of guiding the feature learning by using residual attention learning for improving the recognition performance. Another example is the attention-focused CNN (RA-CNN)~\cite{fu2017look} based on the Attention Proposal Network (APN), which actively identifies the effective region and uses bi-linear interpolation to adjust the scale, and then the enlarged region of interest is used for improved fine-grained classification. %
However, the above works apply the attention mechanism in the 2D domain. Our work extends the attention mechanism to the 3D domain to extract multiscale deformation features based on an effective shape representation~\cite{gao2017sparse}.

\section{Deformation Representation and  Convolution Operator}\label{sec:feature}
\YL{The input of our overall network is based on the recently proposed as-consistent-as-possible (ACAP) deformation representation~\cite{gao2017sparse}, which can cope with large-scale deformations of shapes and is defined only on vertices, making mesh-based convolutions easier to implement. 
To validate it is a good choice, we compare this with a recently proposed general-purpose mesh autoencoder (AE) architecture DEMEA~\cite{Tretschk2019arXiv}. For fair comparison, we compare with DEMEA~\cite{Tretschk2019arXiv} on the COMA~\cite{COMA:ECCV18} dataset with the same setting as used in~\cite{Tretschk2019arXiv}. We also use the same training set and test set split: the training set contains 10 various expressions and 17794 meshes in total. The test set contains 2 challenging expressions (high smile and mouth extreme), with 2671 meshes in total. We set the same latent space dimension (32) and our network only uses a single level of AE to make the network architecture comparable. The results on the test set in terms of average per-vertex errors show that our base AE network with 0.822mm error outperforms DEMEA with 1.05mm error. Such benefits can be more substantial when shapes undergo more substantial deformations, such as various human body poses, so we build our network architecture based on this.}

For a given shape set with $N$ shapes that share the same connectivity each with $V$ vertices, without loss of generality, we choose the first shape as the reference shape. \hyh{For the patch which consists of the $i^{\rm th}$ vertex and its 1-ring neighbor vertices, we can calculate its deformation gradient $\mathbf{T}_{m,i}\in\mathbb{R}^{3\times3}$. The deformation gradient of the patch is defined on the $i^{\rm th}$ vertex of the $m^{\rm th}$ shape, which describes the local deformation w.r.t. the reference shape.}
$\mathbf{T}_{m,i}$ of shape $m$ is obtained by minimizing the following formula:
$
\mathop{\arg\min}_{\mathbf{T}_{m,i}} \sum_{j\in \mathcal{N}_i}{c_{ij}\|(\mathbf{p}_{m,i}-\mathbf{p}_{m,j})-\mathbf{T}_{m,i}(\mathbf{p}_{1,i}-\mathbf{p}_{1,j})\|_2^2}
$
where $\mathcal{N}_i$ is the 1-ring neighbor vertices of  vertex $i$ and $c_{ij} = \alpha_{ij} + \beta_{ij}$ is the cotangent weight~\cite{ARAP_modeling:2007}, where $\alpha_{ij}$ and $\beta_{ij}$ are the angles in the two faces that share a common edge $(i,j)$. 
Then $\mathbf{T}_{m,i}$ can be decomposed as $\mathbf{T}_{m,i} = \mathbf{R}_{m,i}\mathbf{S}_{m,i}$ using the polar decomposition, where $\mathbf{S}_{m,i}\in\mathbb{R}^{3\times3}$ is a symmetry matrix that describes the scaling/shear deformation and $\mathbf{R}_{m,i}\in\mathbb{R}^{3\times3}$ is an orthogonal matrix that describes the rotation. For the rotation matrix $\mathbf{R}_{m,i}$, it can be represented by the rotation axis $\omega_{m,i}$ and rotation angle $\omega_{m,j}$. But the mapping from the rotation matrix to rotation axis and angle is one-to-many. For shapes with large-scale rotations, the rotation axis and rotation angle of adjacency vertices may become inconsistent, which results in artifacts when synthesizing new shapes, as shown in~\cite{gao2017sparse}.
Gao et al. propose a two-step integer optimization to solve the problem which makes the rotation angle and rotation axis of adjacent vertices as consistent as possible. For the details, please refer to~\cite{gao2017sparse}.

Next, for each vertex $i$ on the $m^{\rm th}$ shape, we can obtain the feature $q_{m,i} = \{r_{m,i}, s_{m,i}\}\in \mathbb{R}^9$ by extracting non-trivial elements $r_{m,i} \in \mathbb{R}^3$ and $s_{m,i} \in \mathbb{R}^6$ from the logarithm of rotation matrix $\mathbf{R}_{m,i}$ and \gl{scaling/shear }%
 matrix $\mathbf{S}_{m,i}$ respectively. Finally, the ACAP feature of the $m^{\rm th}$ shape  can be represented by $\{q_{m,i}| 1 \le i \le V\}$.
Due to the usage of the $tanh$ activation function ~\cite{sparsevae2017}, we further linearly scale each element in $r_{m,i}$ and $s_{m,i}$ to  $[-0.95, 0.95]$ separately.
Then we concatenate $q_{m,i}, 1 \le i \le V$ together in the vertex order to form a long vector $X_{m} \in \mathbb{R}^{\mu V \times 1}$ as the feature of the $m^{\rm th}$ shape, where  $\mu=9$ is the dimension of the ACAP feature of each vertex.

We further introduce the graph convolutional operator used in our architecture. As illustrated  in~\cite{sparsevae2017}, the output of the convolution operator for every vertex is a linear combination of the inputs of the vertex and its 1-ring neighbor vertices, along with a bias. The output $y_i$ for the $i^{\rm th}$ vertex is defined as follows:
\begin{equation}
\mathbf{y}_i = \mathbf{W}_{point}\mathbf{x}_i + \mathbf{W}_{neighbor}\frac{1}{D_i}\sum_{j=1}^{D_i}\mathbf{x}_{n_{ij}} + \mathbf{b}
\end{equation}
where $\mathbf{x}_i$ is the input feature vector of the $i^{\rm th}$ vertex, $D_i$ is the degree of the $i^{\rm th}$ vertex, and $n_{ij}$ $(1 \leq j \leq D_i)$ is the $j^{\rm th}$ neighbor vertex of the $i^{\rm th}$ vertex. $\mathbf{W}_{point}$, $\mathbf{W}_{neighbor} \in \mathbb{R}^{\mu \times \mu}$ and $\mathbf{b} \in \mathbb{R}^{\mu}$ are the trainable parameters of the graph convolutional layer. 
\YL{All these weights are shared by all the vertex and their neighborhoods in the same convolutional layer and learned during the training of the network.}

\section{METHODOLOGY}\label{sec:network}

In this section, we introduce our network architecture from these three main aspects: our novel autoencoder structure, attention mechanism and redundant component removal. Firstly, we introduce the novel autoencoder structure. 
Then we describe our attention mechanism, which is applied to help autoencoders to extract multi-scale deformation components. Lastly, we explain how we remove redundant components, followed by implementation details of our network. %
\gl{This architecture is flexible to support multiple levels of scales. In most cases, two levels of deformation scales are sufficient to represent deformation in the dataset of this paper. Please refer to Sec.~\ref{sec:ae_levels} for details.}

\subsection{Autoencoder Block}%

As Fig.~\ref{fig:network} illustrates, we have achieved the multiscale structure by stacking autoencoder blocks. In the first (coarsest) level, we have one autoencoder $AE_0$, and in the second level, $K$ autoencoders $AE_k$ ($k=1, 2, \dots, K$) are built, each focusing on one local region through an attention mechanism, which will be detailed later. \yjrev{The number of second level AEs is %
\YL{determined by} the dimension of the latent space of $AE_0$.}

\yjrev{For each shape $m, 1 \le m \le N$, we represent it by the pre-processed ACAP feature $X_m \in \mathbb{R}^{V \times \mu}$, as described in Sec.~\ref{sec:feature}. We use an encoder to map the feature to a 128-dimensional latent code and a decoder which reconstructs the shape ACAP feature from a latent code $z$. Both of the encoder and decoder have one mesh-based graph convolutional layer and a fully-connected layer and their network structure is symmetrical, where the learnable parameters of the fully connected layer are defined as $\mathbf{C} \in \mathbb{R}^{K_z \times \mu V}$, $K_z$ is the dimension of the latent space. 
Especially, the fully connected layers of encoder and decoder share same learnable parameter $\mathbf{C}$ without bias.
The latent vector $z$ for all the $N$ shapes form a matrix $\mathbf{Z} \in \mathbb{R}^{N\times K_z}$.}
\yjrev{Similar to~\cite{sparsevae2017}, all the layers use the $tanh$ activation function. Figure~\ref{fig:network} illustrates the autoencoder architecture in the left top corner. }

The output $\widehat{X} \in \mathbb{R}^{\mu V}$ of the whole autoencoder block can be scaled back to the ACAP deformation representation and reconstruct the Euclidean coordinates using~\cite{gao2017sparse}. For every autoencoder, we optimize the following loss function that includes three terms: reconstruction loss
that ensures accurate reconstruction of the input, 
sparsity loss $\Omega(\mathbf{C})$ that promotes localized deformation components, and  non-trivial regularization term $\mathcal{V}(\mathbf{Z})$ to avoid creating trivial solutions. 
The total loss for an autoencoder block $AE_k$ is as follows:
\begin{equation}
\mathcal{L}_{AE_k} = \lambda_1L_{recon} + \lambda_2\Omega(\mathbf{C})+ \mathcal{V}(\mathbf{Z})
\end{equation}
where $AE_k$, $0\le k \le K$ represent the $k^{\rm th}$ autoenoder, $\lambda_1, \lambda_2$ are the balancing weights.

The reconstruction loss is \gl{the} MSE (mean square error) loss, defined as $L_{recon}=\frac{1}{N}\sum_{i=1}^N\|X_i-\widehat{X}_i\|_2^2$. For the non-trivial regularization term, $\mathcal{V}(\mathbf{Z})=\frac{1}{K_z}\sum_{j=1}^{K_z}\max((\max_{m}|Z_{jm}|-\theta),0)$, where $Z_{jm}$ is the weight for the $j^{\rm th}$ dimension of the $m^{\rm th}$ shape, and $\theta$ is a positive number and we set $\theta = 5$ in our experiments.

\yjrev{The above two loss terms are \hyh{the} same as the previous work~\cite{sparsevae2017}. But for the loss $\Omega(\mathbf{C})$, there are some difference: }
we choose the step function $\Lambda$ defined above to map geodesic distances to $\{0,1\}$, rather than the previously used clipped linear interpolation function. 
This is because our network architecture 
extracts hierarchical deformation components, so at any level, a fixed component size (rather than a range) is preferred.
Autoencoder blocks at different levels will produce localized deformation components of different scales by adjusting the tunable parameter $d$. Our sparsity loss term $\Omega(\mathbf{C})$ is defined as: $\Omega(\mathbf{C}) = \frac{1}{K_z}\sum_{k=1}^{K_z}\sum_{i=1}^V\Lambda_{ik}\|C_{k,i}\|_2$, where $C_{k,i}$ is the $\mu$-dimensional vector of component $k$ of vertex $i$, $\left\| \cdot \right\|_2$ is group sparsity ($\ell_{2,1}$ norm) , and $\Lambda_{ik}$ is sparsity regularization parameters defined as follows:

\begin{equation}\label{eqn:sparsity}
\Lambda_{ik}=
\left\{
\begin{array}{ll}
0 & d_{ik} < d \\
1 & d_{ik} \geq d
\end{array}
\right.
\end{equation}
where the $\Lambda_{ik}$ is a binary function, where $d_{ik}$ denotes the normalized geodesic distance~\cite{Crane:2013:GH} from vertex $i$ to the center point $c_k$ of component $k$, which is defined as $c_k = \argmax_i\|C_{k,i}\|_2$, and is  updated in each iteration of network training. \yjtvcg{$d$ is a tunable parameter, which controls the size of the deformation region of a component. Larger $d$ corresponds to bigger deformed regions of the shape. For our task, AEs of different levels choose different values of $d$. Please refer to Sec.~\ref{sec:details} for the default value of $d$. 
}

We train the whole network end-to-end by adding all the losses of autoencoder blocks together as $L_{total} = \sum_{k=0}^{K} \mathcal{L}_{AE_k}$, which includes a first-level AE ($AE_0$) and $K$ second-level AEs ($AE_k$, $1\leq k\leq K$).

\subsection{Attention Mechanism for Multiscale Analysis}\label{sec:attention}%

\yjrev{Similar to 2D images, there are many tasks (image recognition) that focus on the different level of images by attention mechanism, such as~\cite{fu2017look}. So we want to explore the 3D shape with multiscale fashion by the attention mechanism. 
\hyh{We design an attention mechanism that facilitates our autoencoder blocks extracting the multiscale deformations components.}}

\gl{Mostly, the deformation datasets
(e.g. human, horse and fabric) have both global scale and local scale \hyh{deformations}. Hence, it is naturally to extract these deformations \hyh{in} multiscale manner.}
\yjtvcg{To make the second level AEs focus on sub-regions to extract finer level components and then form a multiscale structure, we extract learnable attention masks from the fully-connected layer of the first-level AE $AE_0$.}
Our attention mechanism is shown in the bottom left corner of Fig.~\ref{fig:network}.
Due to the sparsity constraint $\Omega(\mathbf{C})$, the parameter $\mathbf{C}$ of the fully connected layer represents the sparse deformation components. 
\yjrev{The $C_k \in \mathcal{R}^{V\mu \times 1}, 1 \leq k \leq K$ that is denoted as each vector in $\mathbf{C}$ represents a deformed sub-region of the shape. The deformed sub-regions can be regarded as the interested mask of the second level AEs. So in every iteration of training, we can extract each row of $\mathbf{C}$ by setting the latent vector to a one-hot vector for second level AEs:}

\begin{equation}
C_k = \mathbf{C}^T \times OH_k
\end{equation}
where the $OH_k$ is a $K$-dim column vector and $k^{\rm th}$ entry of $OH_k$ is $1$ and the rest are $0$. \yj{Then we reshape $C_k$ to a 2D array with the size $V \times \mu$, denoted as $C_{k}^{r}$. The unnormalized attention mask $am_{k,i} \in \mathbb{R}^{K \times V}$ for the $k^{\rm th}$ component of the $i^{\rm th}$ vertex is defined as
\begin{equation}
am_{k,i}=\sum_{j=1}^{\mu}{C_{k, ij}^{r}}^2.
\end{equation}
where the $C_{k, ij}^{r}$ is the $(i,j)$ entry of $C_{k}^{r}$.
We further normalize it to obtain the normalized attention mask $AM \in \mathbb{R}^{K \times V}$, where
\begin{equation}
AM_{k,i} = \frac{am_{k,i}}{\sum_{k=1}^K{am_{k,i}}}.
\label{eqn:1}
\end{equation}

So for the first-level autoencoder $AE_0$, the residual value of the reconstruction is $X - \widehat{X}$ and the normalized attention mask is $AM$. We reshape $(X-\widehat{X})$ to a 2D array $X_{res} \in \mathbb{R}^{V \times \mu}$.
For the second-level autoencoder $AE_k, 1 \le k \le K$ as in Fig.~\ref{fig:network}, its input is $diag(AM_k) \times X_{res}$, where $diag(\cdot)$ returns a square diagonal matrix with the elements of the vector on the main diagonal. \yjtvcg{The input of each second-level AE is therefore the weighted (by the corresponding attention mask) residual of the first-level AE.} Therefore, this attention mechanism ensures that the second-level AEs can reconstruct smaller scale deformations that cannot be well captured by $AE_0$, and each $AE_k$ focuses on an individual local region. 
The sum of every column of $AM$ is one according to Eq.~\ref{eqn:1}, which  ensures 
the sum of inputs to $AE_k$, $1 \leq k \leq K$ equals the residual value $X_{res}$ of the first-level autoencoder $AE_0$.

}

Under the supervision of loss function and attention mechanism, \yjrev{the first level autoencoder $AE_0$ is capable of capturing the large scale deformation \hyh{and the second level AEs can capture the smaller scale deformations on specified regions of the shape. Consequently,}} our network can learn the multiscale deformation components of the whole shape set, \yjrev{the multiscale deformation components can be extracted from the \hyh{parameters} of fully connected layers of all AEs}. %

\subsection{Redundant Component Removal}\label{sec:removal}
\begin{figure*}[ht]
\centering
\includegraphics[width=0.99\linewidth]{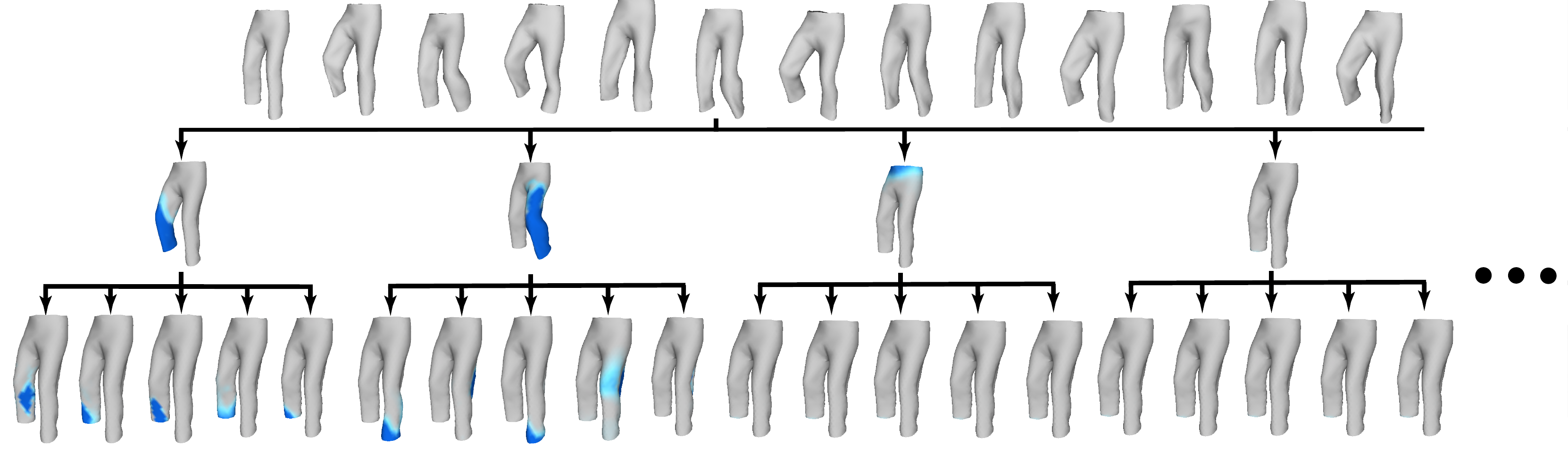} 

\caption{The multiscale structure of deformation components on the Pants dataset~\cite{white2007siggraph}. In the figure, we don't filter the redundant components, the symbol `$\cdots$' represents there are the similar results which contains slight deformation. This result is comparable with Fig.~\ref{fig:pantstree} which is processed by the Sec4.3. As a result, our method can learn deformation components of different scales. In the deformation components, there will be some finite component to represent the whole dataset, many redundant components which contains some slight deformations will appear.}
\label{fig:pants_treeall}
\end{figure*}

\begin{figure}[ht]

\centering
\includegraphics[width=0.95\linewidth]{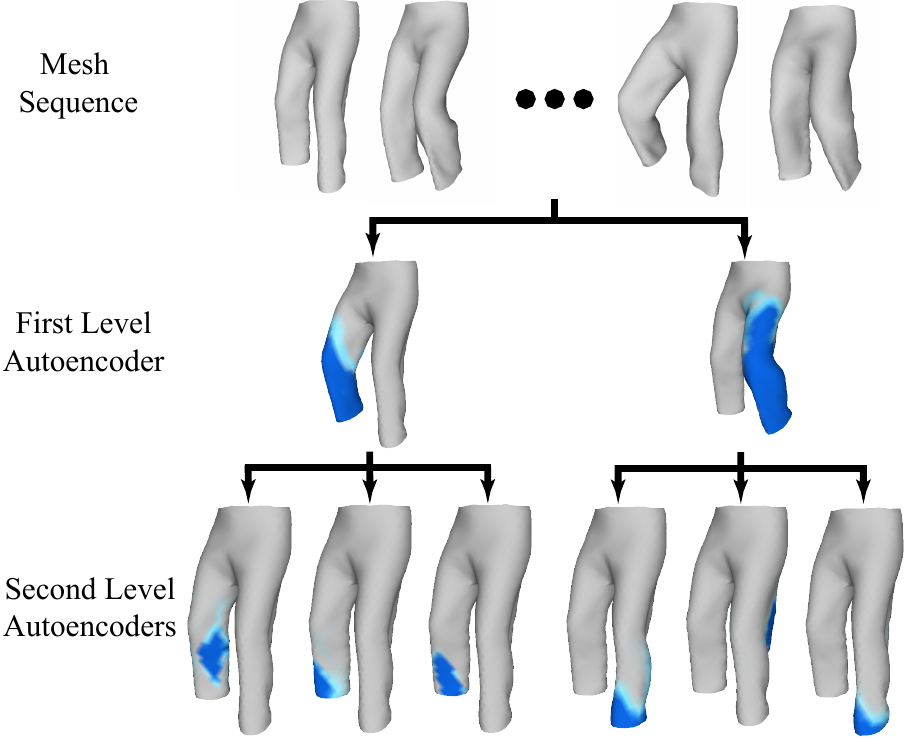} %

\caption{The multiscale structure of deformation components on the running pants data set~\cite{white2007siggraph}. In the figure, we have removed redundant components. Our method can learn  deformation components at different scales. The first row shows example shapes of the running pants, the second row gives deformation components caused by leg movement, which are extracted by the first-level AE, and the third row presents the detailed deformations of the cloth from the second-level AEs.}%
\label{fig:pantstree}
\end{figure}

\yjrev{For all autoencoder blocks, we extract the fix number of deformation components for each AE. For fair comparison and capture all deformations, we set $K_z=10$ for $AE_0$, $K_z=5$ for $AE_k, 1\le k \le K$.}
\yjrev{Because the multiscale analysis and our setting is to extract as much as possible to avoid missing any components, so there are some redundant components in second level AEs, \hyh{which means the components does not contain any deformation, or only slight deformations compared to the reference mesh.} So we remove these components if the contained information is lower than the given threshold value. All results in our paper are processed by the redundant component removal. \hyh{The Fig. ~\ref{fig:pants_treeall} shows the results of the output of network without the process (there are some components with slight deformations compared to the reference mesh). Fig.~\ref{fig:pantstree} illustrates some finite components are sufficient enough to represent the whole dataset during the deformation components analysis. Thus, we will remove the redundant components which contain slight deformations defined in Eq.~\ref{eq:remove}.}

The process aims to make our results more brief and reasonable and is done after network training. It can gain trade off between the multiscale decomposition and avoiding overfitting on training data. 
}
\yj{
As a result, we could exclude these redundant deformation components. We define the following deformation strength of deformation components to filter slight or noisy deformation components. The strength $I$ on the features $X_m$ is defined as:
\begin{equation}\label{eq:remove}
I(X_m)=\frac{
\sum_{i=1}^{V}{1(\|(X_{diff})_i\|>\epsilon_1) \|(X_{diff})_i\|}
}{\sum_{i=1}^V{1(\|(X_{diff})_i\|>\epsilon_1)}}
\end{equation}
where $X_{diff} = X_m - X_r$, and $X_r, X_m \in \mathbb{R}^{V \times \mu}$ are the features of the reference mesh and the extracted deformation components from the autoencoders respectively.
$\|\cdot\|$ is the $\ell_2$ norm of the vector, and $1(\cdot)$ gives 1 if the condition is true, and 0 otherwise. $\epsilon_1 = 1e-6$.}

\begin{figure}[ht]
\centering
\includegraphics[width=0.75\linewidth]{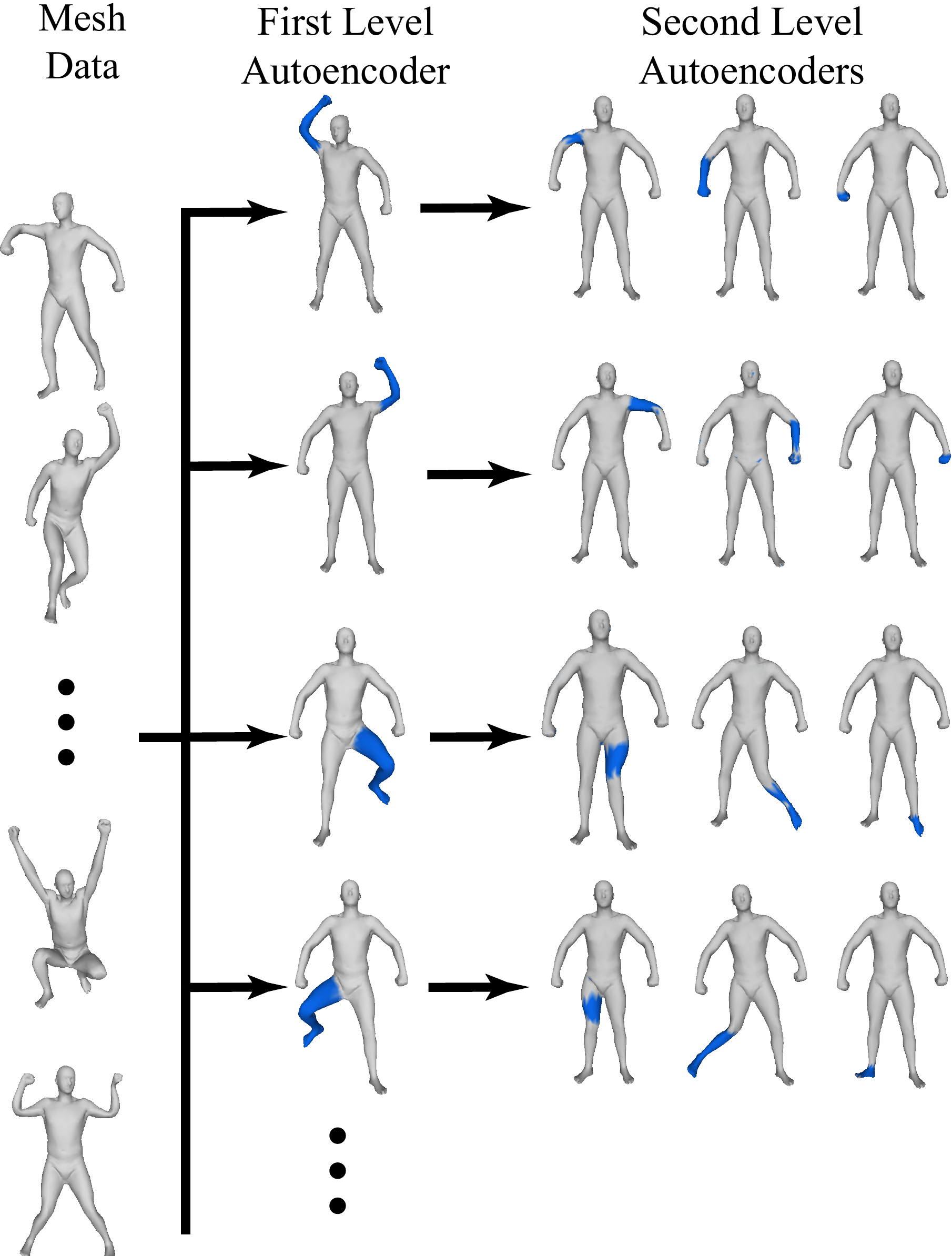} 

\caption{The multiscale structure of deformation components on the shape set SCAPE~\cite{anguelov2005scape}. In the figure, we filter the redundant components. As a result, our method can learn deformation components of different scales. The first column shows some examples from the SCAPE dataset, the second column presents coarse level deformation components from the first-level $AE_0$, and the third column gives the fine level deformation components from the second-level $AE_k, 1 \le k\le K$.}%
\label{fig:scapetree}
\end{figure}

If the extracted deformation component \hyh{corresponds to a} slight deformation, defined as its strength being smaller than a threshold $\epsilon_2$, we will remove the component.
In our experiments, we set $\epsilon_2 = 1e-2$. Finally, we obtain the multiscale structure of the deformation components, as shown in Figs.~\ref{fig:scapetree}, \ref{fig:dresstree}, \ref{fig:fattree}, \ref{fig:pantstree}, \ref{fig:swingtree} and \ref{fig:flagtree}.
\yjtvcg{, where deformations at different scales are indicated with arrows.}

\begin{figure}[ht]

\begin{center}
  \includegraphics[width=0.7\linewidth]{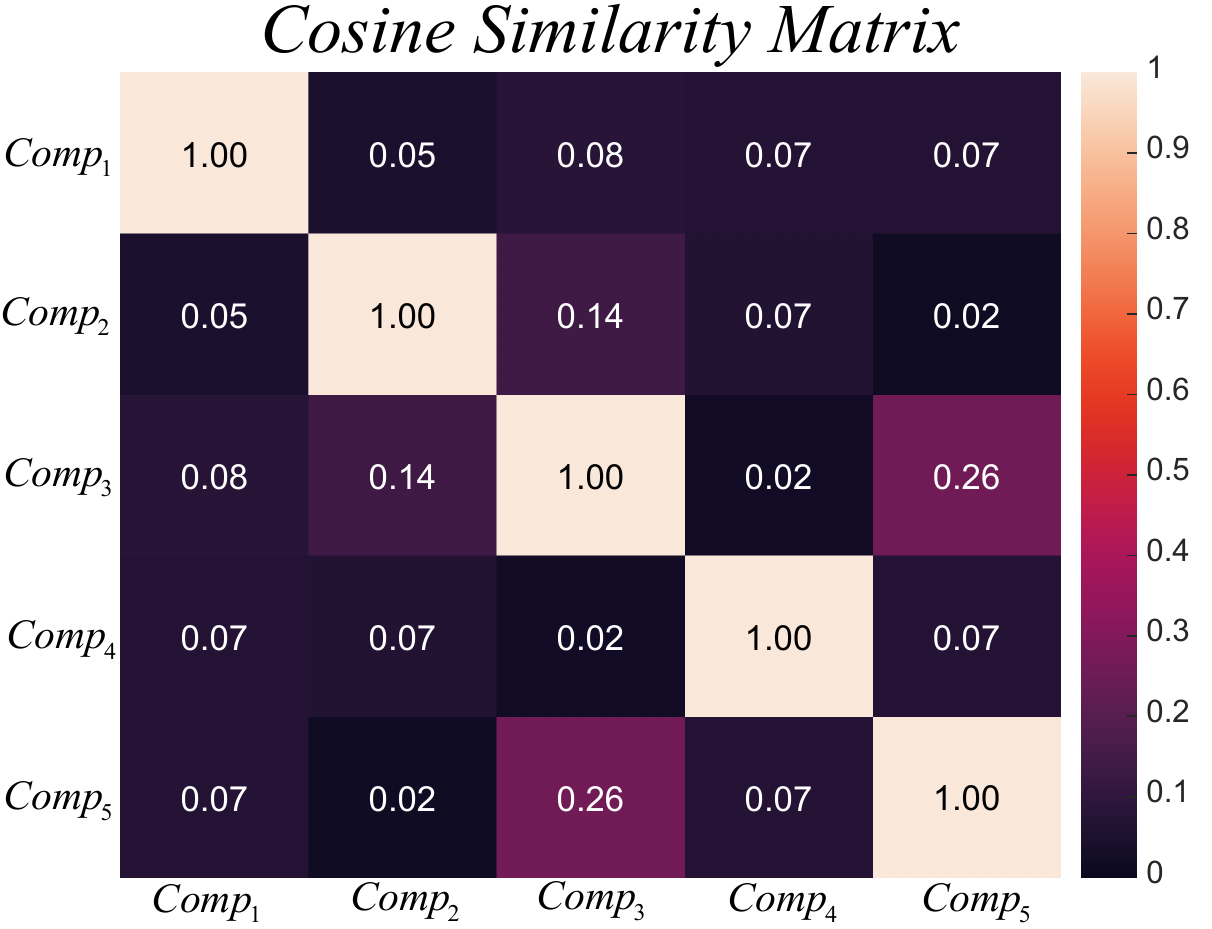}
\end{center}

  \caption{\yjtvcg{Visualization of the cosine similarity matrix of the components extracted from the first-level AE\hyh{. It} shows that the components have low similarity. The value (0-1) in each grid indicates the similarity between two components. Larger values mean more similar.}}

\label{fig:sim}
\end{figure}

\yjtvcg{We also check if our network produces similar components (near duplicates). To achieve this,  we test the similarity of the extracted deformation components.
Fig.~\ref{fig:sim} visualizes the cosine similarity matrix of the components extracted from the first-level AE.
It shows that the components have low similarity, because our AE applies the localization constraint and reconstruction error minimization to ensure different \hyh{components} in the latent space represent different parts on the shape; having duplicated components would lead to reduced representation capability, so higher total loss.}

\begin{figure}[ht]
\begin{center}
\includegraphics[width=0.75\linewidth]{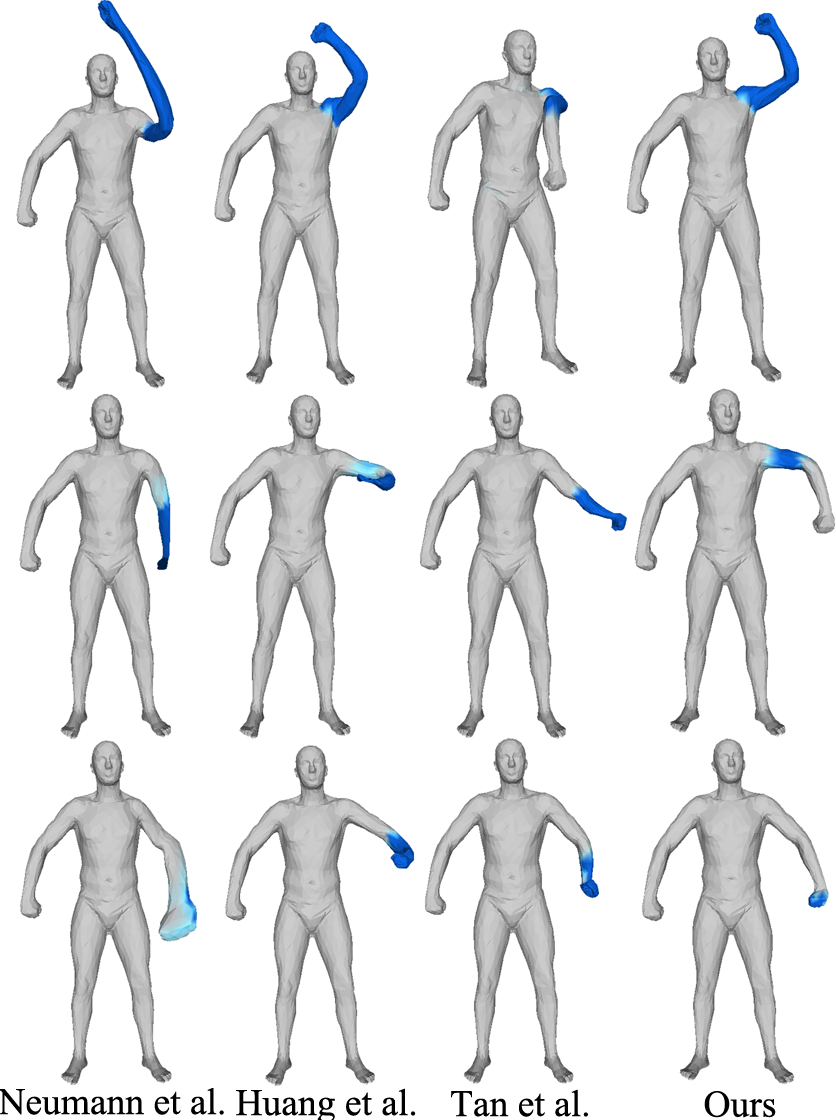}
\end{center}
\caption{Comparison of deformation components located in the left arm of SCAPE~\cite{anguelov2005scape}, which are extracted by different methods. The deformed region is highlighted in blue. Every row shows the components located in a similar region. It shows that our results are more reasonable.}
\label{fig:leftarm}%
\end{figure}

\subsection{Implementation Details}\label{sec:details}

Our experiments were carried out on a computer with an i7-6850K CPU, 16GB RAM and an Nvidia GTX 1080Ti GPU.

\noindent\textbf{Datasets:} \yj{We compare with the state-of-the-art methods on the SCAPE dataset~\cite{anguelov2005scape}, Horse dataset~\cite{sumner2004deformation}, Face dataset~\cite{zhang-siggraph2004-stfaces}, Humanoid dataset~\cite{wang2017articulated}, Dress dataset, Pants dataset~\cite{white2007siggraph}, \yjtvcg{Flag dataset, Skirt Dataset, Fat person   (ID:50002) from the Dyna~\cite{Dyna:SIGGRAPH:2015} Dataset, Coma~\cite{COMA:ECCV18} Dataset}, Swing and Jumping datasets~\cite{Vlasic2008}. The Dress, Flag and Skirt datasets were synthesized by the NVIDIA Clothing Tools on 3ds MAX. \yjrev{Our used data range from rigged deformation like human motion to non-rigged deformation (Face, cloth), from small dataset (several hundreds shapes) to large dataset (thousands of shapes(Dyna)). For the above datasets, they are mainly sequence models which have similar deformation between two neighbour shapes. So for our model's generaliablity, we select one from ten models to training and the rest to testing, the ratio is 9:1 for test and training. For un-sequence data like SCAPE, we spilt the training set and test set randomly with a ratio 1:1. \yjrev{A special case for Coma dataset, we use same setting as the DEMEA~\cite{Tretschk2019arXiv} for fair comparison.} The statistics of the dataset are shown in Table~\ref{tab:statdata}. The table lists the number of shapes that dataset contains, training examples and testing examples. All data is very easy to obtain, some are the public data, some are the synthesized data by the professional software. We will release the synthesized data for future research.
}}

\begin{table}[htbp]
  \centering
  \caption{Data Statistics. We summarize the data statistics of 10 datasets in our experiments. Each data is split into training set and test set with a ratio of 1:9. For the Coma~\cite{COMA:ECCV18} dataset, we use the same setting with DEMEA~\cite{Tretschk2019arXiv} for fair comparison.}
    \begin{tabular}{cccc}
    \toprule
    DataSet & \# All Shapes & \# Training Shapes & \#Testing Shapes \\
    \midrule
    Swing & 280   & 28    & 252 \\
    Scape & 71    & 36    & 35 \\
    Pants & 241   & 25    & 216 \\
    Humanmoid & 154   & 16    & 138 \\
    Horse & 49    & 5     & 44 \\
    Flag  & 500   & 50    & 450 \\
    Dress & 500   & 50    & 450 \\
    Jump  & 150   & 15    & 135 \\
    Fat   & 4737  & 469   & 4274 \\
    Face  & 385   & 39    & 346 \\
    Skirt & 231   & 23    & 208 \\
    Coma  & 20465 & 17794 & 2671 \\
    \bottomrule
    \end{tabular}%
  \label{tab:statdata}%
\end{table}%

\yjrev{In our experiments, our network takes the ACAP features of 3D shapes as input, which can describe the large scale deformations and be calculated by the novel work~\cite{gao2017sparse}.} We have two levels of autoencoders, the first-level only has one AE ($AE_0$), and the second-level has the same number of AEs ($AE_k,1 \le k \le K$) as the attention masks. Since in most wild data sets, the deformation of the shape is not very exaggerated, two levels of autoencoders are enough to extract the multiscale localized deformation components in our experiments, but this can be extended if necessary. 
\yjrev{For above different categories, we use fixed hyperparameters. We perform experiments on the SCAPE dataset to demonstrate how we choose the suitable hyper-parameter in the Sec.~\ref{sec:ablation}.
As shown in that section, our stacked AEs has the lowest error with the following default parameters: }
$\lambda_1=10.0, \lambda_2=1.0$, where $\lambda_1, \lambda_2$ are the weights of reconstruction error and sparsity constrain term, respectively. $d=[d_1, d_2]=[0.4, 0.2], K_z=[10, 5]$, corresponding to the coarse and fine levels. 
\yjrev{Here, we train the whole network end-to-end rather than in a separate manner. We set the learning rate as $0.001$ with the exponential decay by the ADAM solver~\cite{Kingma2015} to train the network end-to-end until it converges, which takes approximately $10,000$ epochs.}
\yjrev{For all AEs, we set the batch size as 256, which is randomly sampled form the training data set. For a typical dataset, the training of stacked AEs takes about 10 hours. Once our network is trained, the extracted components will output efficiently: outputing one component takes only about 50 milliseconds.}

As we will later discuss, compared with separate training in Table~\ref{tab:eval_train_att}, training the network end-to-end can result in smaller reconstruction errors $E_{rms}$ on all data sets.
The main reason is that the network can adjust the attention mask by minimizing the loss function, and conversely, the adjusted attention mask will result in smaller reconstruction errors $E_{rms}$. Under the collaboration, we get better results as shown in Table~\ref{tab:eval_train_att}.

\section{Experimental Results \& Evaluation}\label{sec:result}
In this section, we will evaluate our method on the above datasets from the following aspects: Quantitative Evaluation, Qualitative Evaluation and Applications.

\begin{table*}
\fontsize{9}{12}\selectfont
\tabcaption{Errors of applying our method to generate unseen data from Horse \cite{Sumner:2004:DTT:1015706.1015736}, Face~\cite{zhang-siggraph2004-stfaces}, Humanoid~\cite{wang2017articulated}, Pants~\cite{white2007siggraph}, \yjtvcg{Flag dataset, Fat person (ID:50002 from the Dyna dataset~\cite{Dyna:SIGGRAPH:2015})}, Swing and Jumping~\cite{Vlasic2008} datasets. From the table, the generation ability of our network is better than the other methods on the $E_{rms}$ error and $STED$ error. 
}
\label{moreevaluation}
\centering
\begin{tabular}{llcccccc}
\toprule
\multirow{2}{*}{Dataset}&\multirow{2}{*}{Metric}&\multicolumn{5}{c}{Method} \\
\cmidrule(r){3-8}
& & Ours  &Tan et al. &Wang et al.&Huang et al.  & Neumann et al. & Bernard et al.\\
\midrule
\multirow{2}{*}{Horse}& $E_{rms}$ & $\mathbf{6.9246}$ & $12.9605$ & $29.6090$ & $18.0624$ & $7.3682$ & $20.1994$\\
\cmidrule(r){2-8}
&$STED$& $\mathbf{0.0336}$ & $0.04004$ & $0.04332$ & $0.05273$ & $0.08074$ & $0.4111$\\
\midrule
\multirow{2}{*}{Face}& $E_{rms}$& $\mathbf{1.4409}$ & $2.9083$ & $8.5620$ & $12.3221$&$2.9106$ & $2.9853$ \\
\cmidrule(r){2-8}
&$STED$& $\mathbf{0.0071}$ & $0.007344$ & $0.01320$ & $0.01827$ & $0.008611$ & $0.02662$ \\
\midrule
\multirow{2}{*}{Jumping}& $E_{rms}$& $\mathbf{16.3475}$ & $24.4827$ & $44.3362$ & $37.9915$&$29.3368$ & $49.9374$\\
\cmidrule(r){2-8}
&$STED$& $\mathbf{0.0321}$ & $0.04862$ & $0.05400$ & $0.06305$ & $0.1268$ & $0.4308$\\
\midrule
\multirow{2}{*}{Humanoid} & $E_{rms}$& $\mathbf{3.2127}$ & $3.4912$ & $60.9925$ & $16.1995$ & $14.3610$ & $6.6320$\\
\cmidrule(r){2-8}
&$STED$& $0.0226$ & $\mathbf{0.01313}$ & $0.03757$ & $0.02247$ & $0.07319$ & $0.04612$\\
\midrule
\multirow{2}{*}{Swing} & $E_{rms}$& $\mathbf{12.2615}$ & $14.0836$ & $29.5329$ & $24.495$ & $15.1942$ & $22.6571$\\
\cmidrule(r){2-8}
&$STED$& $\mathbf{0.0311}$ & $0.03789$ & $0.04224$ & $0.04343$ & $0.0830$ & $0.1139$\\
\midrule
\multirow{2}{*}{Pants} & $E_{rms}$& $\mathbf{6.4083}$ & $7.8986$ & $39.2946$ & $10.1880$ & $28.4118$ & $23.6785$\\
\cmidrule(r){2-8}
&$STED$& $\mathbf{0.0372}$ & $0.0414$ & $0.0540$ & $0.04958$ & $0.1762$ & $0.06484$\\
\midrule
\multirow{2}{*}{Dress}& $E_{rms}$ & $\mathbf{11.5117}$ & $34.0579$ & $35.2816$ & $35.2340$ & $55.5806$ & $12.2239$\\
\cmidrule(r){2-8}
&$STED$& $\mathbf{0.0397}$ & $0.0415$ & $0.0635$ & $0.0782$ & $0.0784$ & $0.2167$\\
\midrule
\multirow{2}{*}{Fat}& $E_{rms}$& $\mathbf{4.3456}$ & $4.5609$ & $25.9187$ & $5.3215$ & $7.4522$ & $5.3348$ \\
\cmidrule(r){2-8}
&$STED$& $0.0052$ & $0.0053$ & $0.0055$ & $\mathbf{0.0035}$ & $0.0372$ & $0.0289$ \\
\midrule
\multirow{2}{*}{Flag}& $E_{rms}$& $\mathbf{20.0627}$ & $26.3174$ & $62.2925$ & $51.2551$&$23.1364$ & $23.1535$\\
\cmidrule(r){2-8}
&$STED$& $\mathbf{0.0157}$ & $0.0176$ & $0.0354$ & $0.2183$ & $0.0914$ & $0.0169$\\
\bottomrule
\end{tabular}

\end{table*}

\subsection{Quantitative Evaluation}%

\begin{figure}[ht]
\begin{center}
\includegraphics[width=0.80\linewidth]{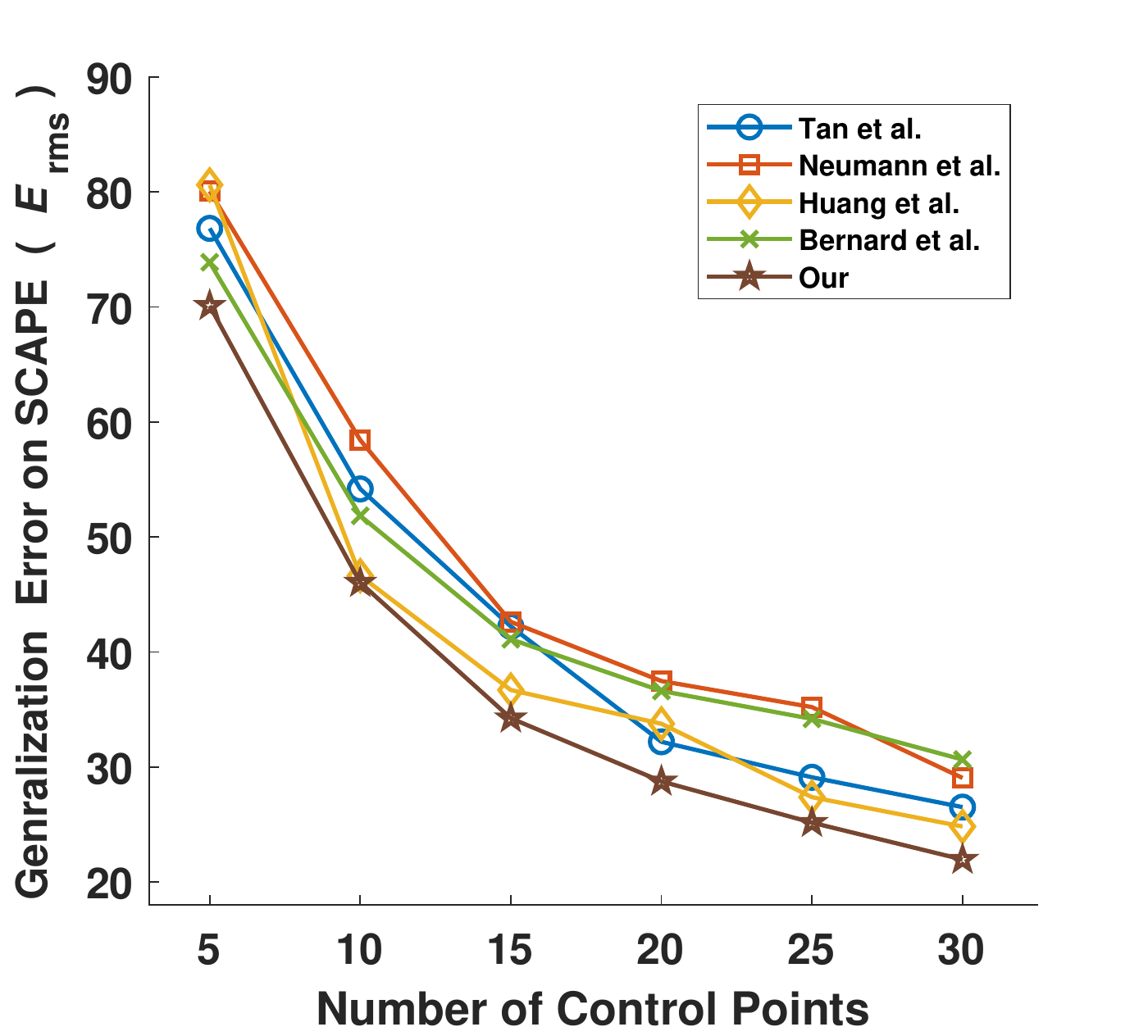}
\end{center}%
\caption{The reconstruction error using sparse control points to deform the SCAPE data set~\cite{anguelov2005scape}. The control points are obtained by furthest point sampling. The generalization error is measured by the data-driven deformation with the extracted deformation components of various methods. In this figure, our performance is better than the other methods with lower errors.}
\label{fig:sparsecontrol}
\end{figure}

We compare the generation ability of our method with the state-of-the-art methods~\cite{bernard2016linear,huang2014sparse,wang2017articulated,sparsevae2017,neumann2013sparse} on  various datasets.
In this experiment, we select one model from every ten models for training and the remaining for testing. After training, we align all the models and scale them into a unit ball. Then we use $E_{rms}$ (root mean square) error~\cite{kavan2010fast} and $STED$ error~\cite{vasa2011perception} to compare the generalization error on the test data (i.e., the reconstruction error for unseen data) with the various methods. \yjtvcg{In particular, $STED$ error is designed for motion sequences with a focus on `perceptual' error of models. }
To ensure fairness, we train each autoencoder to extract 50 components.
As Table~\ref{moreevaluation} shows, the performance of our method is better than the existing methods on the $E_{rms}$ and $STED$. 
\yj{Because the Euclidean coordinate representation is sensitive to rotation, the extracted deformation components of the methods~\cite{neumann2013sparse,bernard2016linear}  have more artifacts and implausible deformation, leading to larger reconstruction errors. Due to the limitation of the edge lengths and dihedral angle representation, the reconstruction using the method~\cite{wang2017articulated} can also be inaccurate and unstable. The method~\cite{huang2014sparse} is not capable of encoding  large-scale deformation (e.g. folds on the fabric), so it cannot recover the original deformation accurately in such cases. The method~\cite{sparsevae2017} uses a large-scale deformation representation to achieve good performance, but it cannot produce multiscale deformation components.
In comparison, our method can keep lower reconstruction errors by using stacked AEs and \hyh{analyze} the residual value of the first-level AE, extracting effective multiscale deformation components.

Meanwhile, our extracted components are served for data-driven deformation. In the real world, the user usually edits the shape by a limited number of control points. To demonstrate the ability of each method to reconstruct deformed models by the limited control points,} we use the furthest point sampling~\cite{kavan2010fast} to sample the control points to ensure that the sampled points distribute on the shape evenly. Under the constraint of the control points, we use the same number of the extracted components to perform data-driven deformation on the SCAPE dataset. As shown in Fig.~\ref{fig:sparsecontrol}, the reconstruction by our extracted components always keep a lower error. In comparison, due to the use of Euclidean coordinate representation, the methods~\cite{bernard2016linear,neumann2013sparse} fail to reconstruct shapes accurately. Our multiscale deformation components better characterize the deformation of the shape, leading to reduced errors.

\begin{figure}[ht]
\begin{center}
\includegraphics[width=0.98\linewidth]{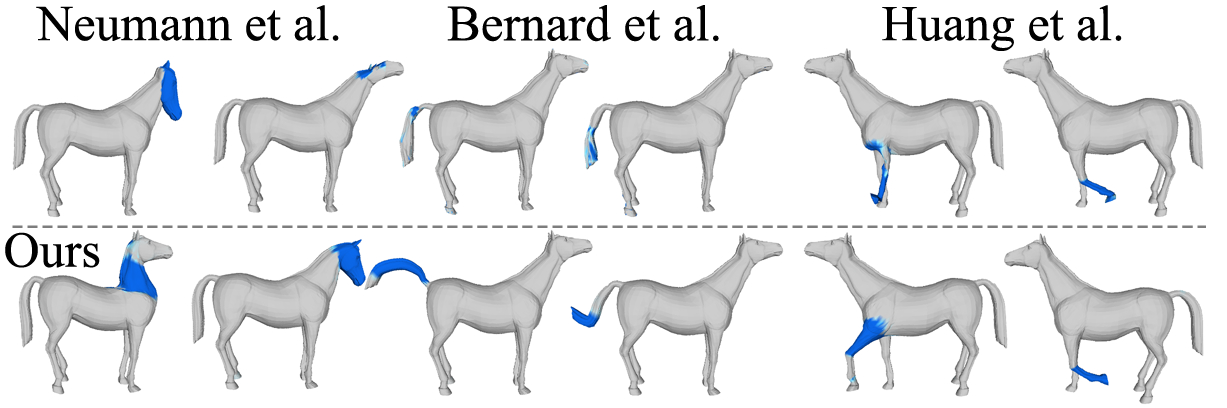}
\end{center}
\caption{Comparison of deformation components located in a similar area on the Horse~\cite{sumner2005mesh}, which are extracted by different methods. The deformed regions are highlighted in blue. The first row shows the results of other methods, and the second row gives the results of our method. Every column shows a component located in a similar region.}
\label{fig:horsecomp}%
\end{figure}

\begin{figure*}[ht]
\centering
\includegraphics[width=0.9\linewidth]{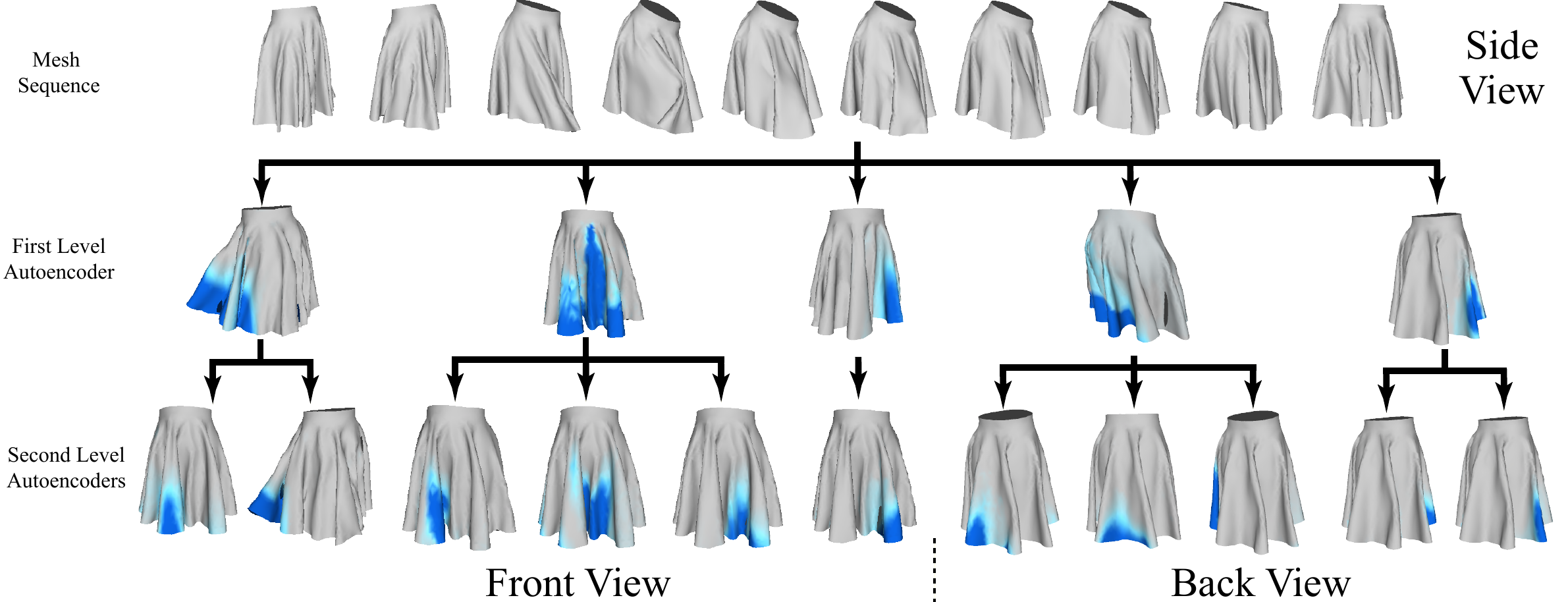} 

\caption{\yjrev{The multiscale structure of deformation components on the skirt cloth dataset, which is our simulated data by physics engine. The data contains more folds of cloth and complex motion. In the figure, we filter the redundant components. The first row is some samples in the skirt dataset. And the second and third rows are the visualization of deformation components extracted from the first level AE and second level AEs respectively. As a result, our method can learn deformation components of different scales with multiscale structure.}}
\label{fig:dresstree2}
\end{figure*}

\begin{figure*}[ht]

\centering
\includegraphics[width=0.99\linewidth]{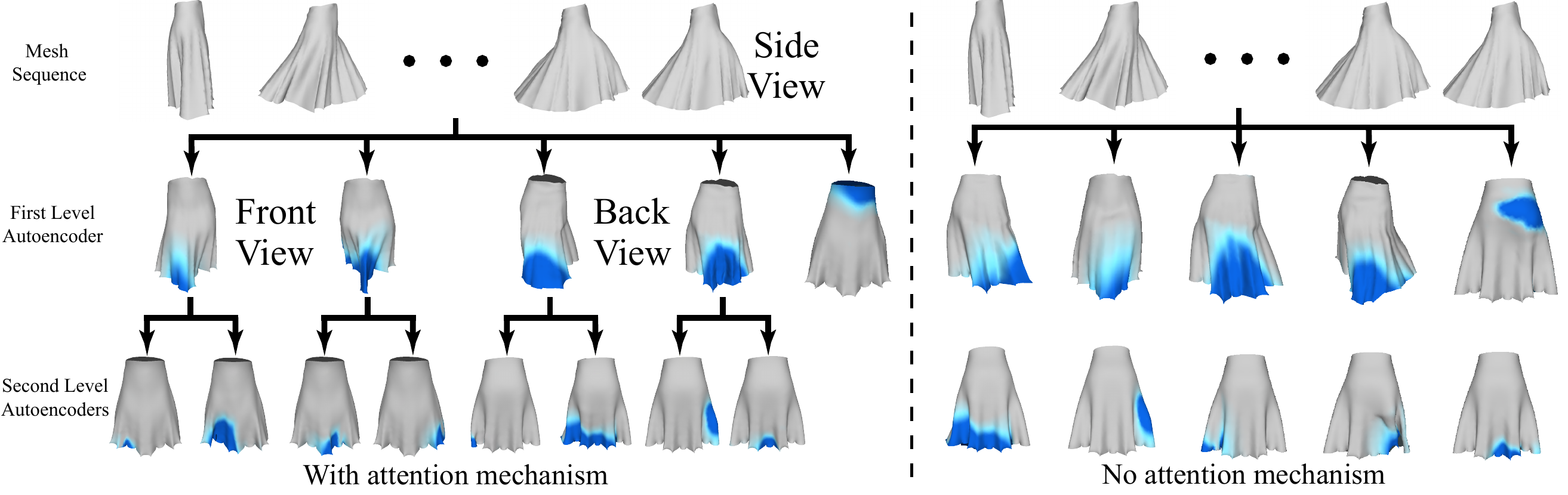} %

\caption{\yjrev{The multiscale structure of deformation components on the Dress data set (a lady walking forward in a skirt). In the figure, we have removed redundant components. In the left part, the result is with the attention mechanism. The first row shows some examples in the dress dataset from the side view. The second row presents the deformation caused by leg movement and the wind, which are extracted by the first-level AE ($AE_0$). The third row is the detailed deformations of the cloth during the movement, which are extracted by the second-level AEs ($AE_k$, $1\le k \le K$). In the second row, the two results on the left are the deformations of the front of the skirt, and the two results on the right are the deformation of the back of the skirt. In the right part, the results is without attention mechanism. It illustrate that the results no longer have a multiscale structure, like the left part shows. The second and third row represent the deformation components that are extracted by first level AE and second level AEs respectively, which is processed by removal of redundant components. The first row is the some sample data in Dress dataset.}}
\label{fig:dresstree}
\end{figure*}

\begin{figure*}[!ht]
  
  \centering
  \includegraphics[width=\linewidth]{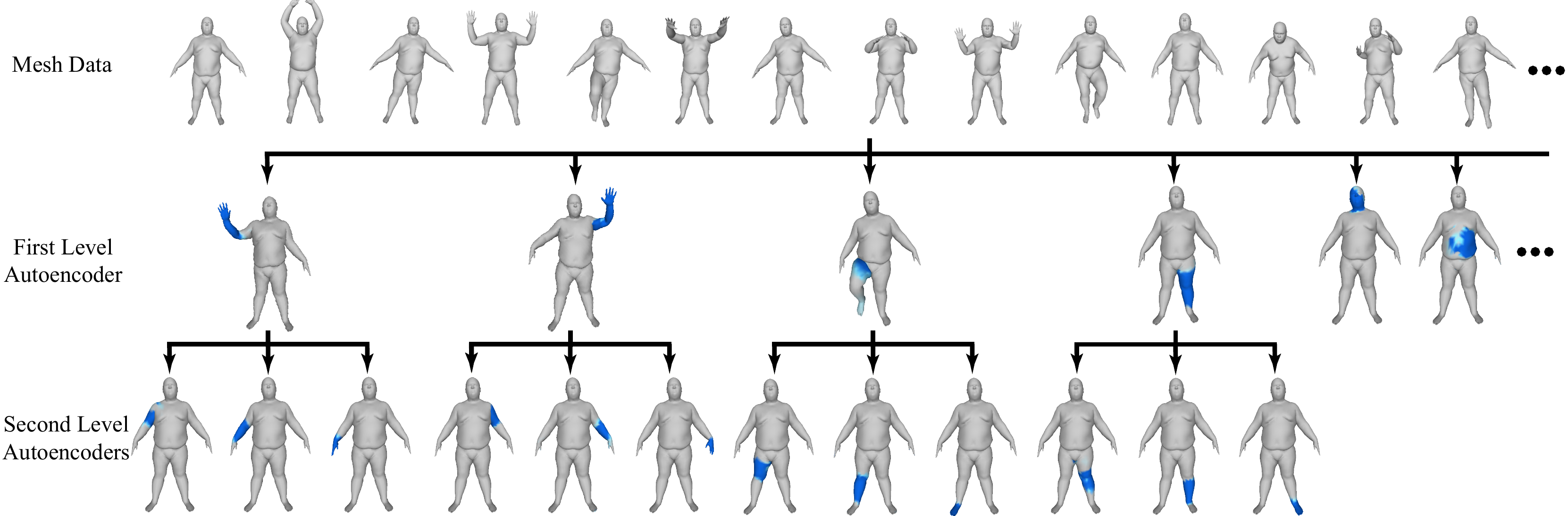}
  
  \caption{\yjtvcg{The multiscale structure of deformation components on the fat person   (ID: 50002) from the Dyna~\cite{Dyna:SIGGRAPH:2015} dataset. The first row shows example shapes in the dataset, the second row presents coarse-level deformation components from the first-level $AE_0$, and the third row shows the fine-level deformation components from the second-level AEs.}}
  \label{fig:fattree}
\end{figure*}

\begin{figure}[!ht]
  
  \centering

  \includegraphics[width=0.95\linewidth]{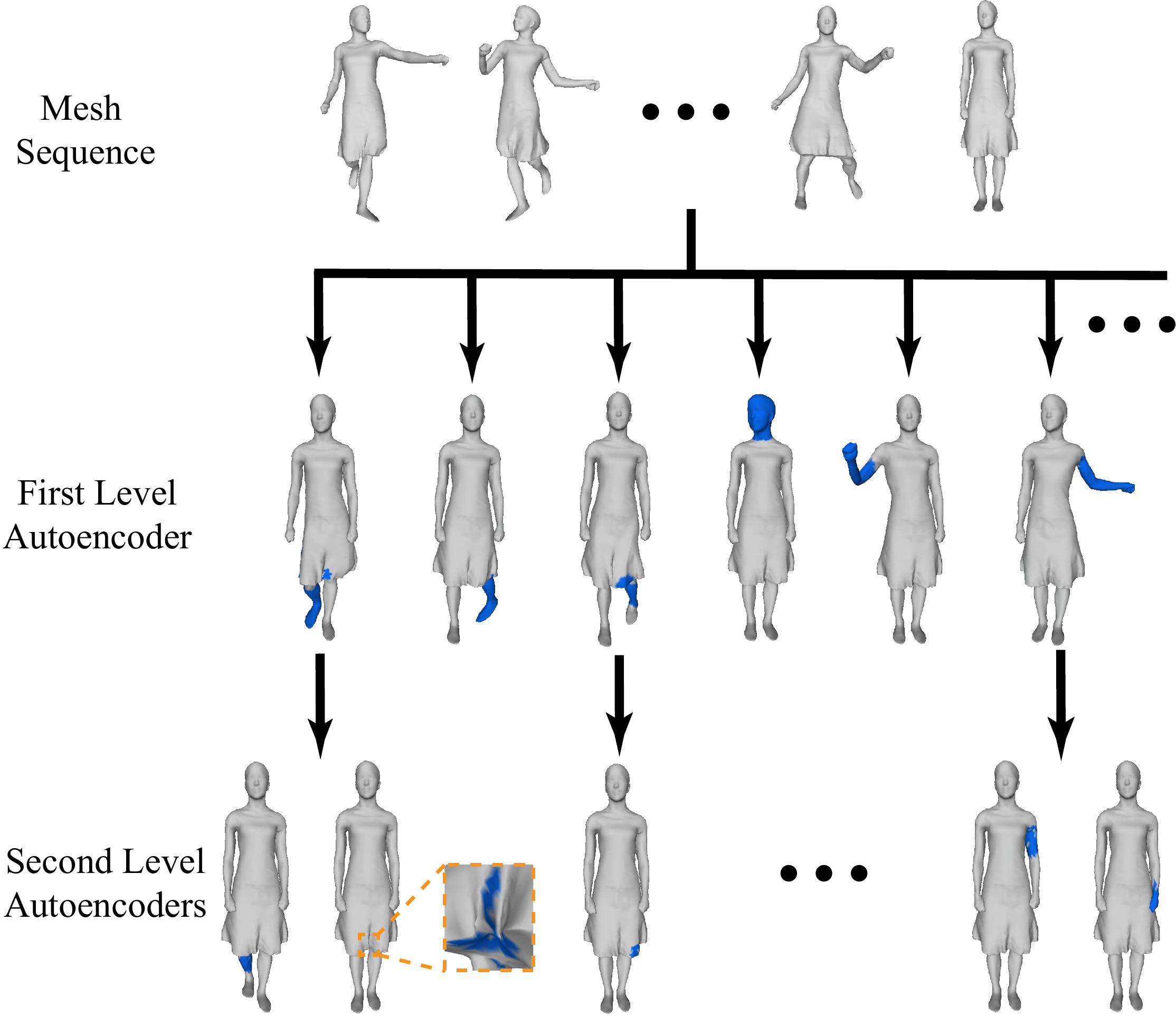}
  
  \caption{\yjtvcg{The multiscale structure of deformation components on the Swing~\cite{Vlasic2008} dataset. The first row shows some example shapes in the dataset, the second row presents coarse-level deformation components from the first-level $AE_0$, and the third row shows the fine-level deformation components from the second-level AEs ($AE_k, 1 \le k\le K$).
  }}
  \label{fig:swingtree}
\end{figure}

\begin{figure}[!ht]
  
  \centering
  \includegraphics[width=0.95\linewidth]{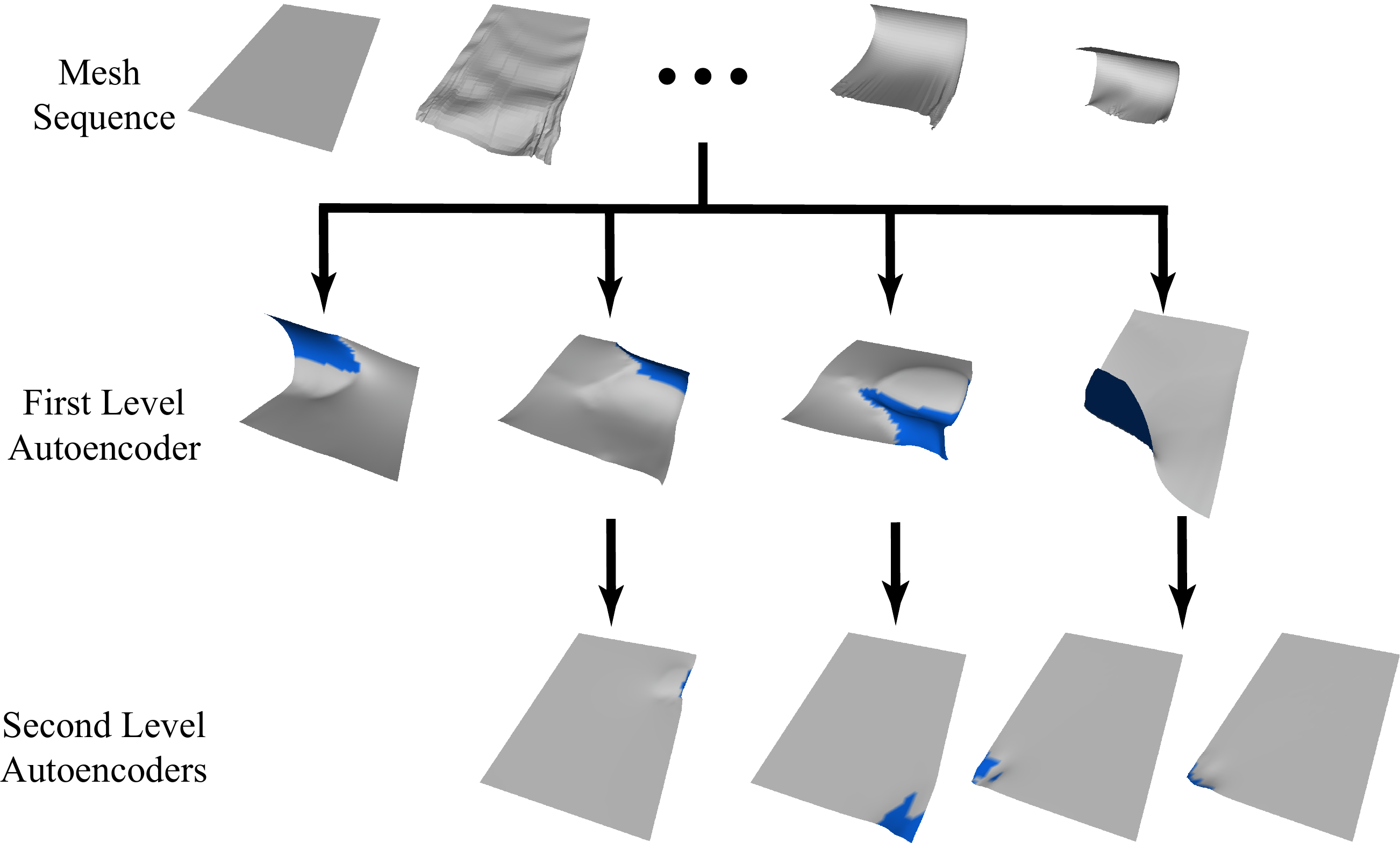}
  
  \caption{\yjtvcg{The multiscale structure of deformation components on the Flag dataset extracted by our method. The first row shows some example shapes of the dataset, the second row presents coarse-level deformation components from the first-level AE ($AE_0$), and the third row shows the fine-level deformation components from the second-level AEs ($AE_k$, $1 \le k\le K$).
  }}
  \label{fig:flagtree}
\end{figure}

\begin{figure}[ht]
\begin{center}
\includegraphics[width=0.80\linewidth]{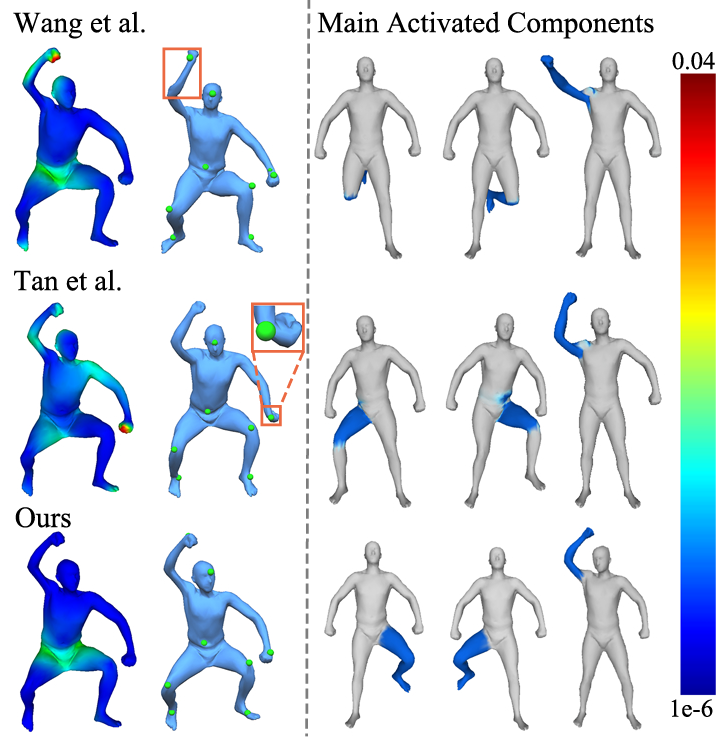}
\end{center}%
\caption{Comparison of shape reconstruction with different methods. \yjtvcg{The first column shows the error heat maps on the ground truth shape between the editing results and ground truth. The second column presents the editing results of different methods. The right side of the figure shows the three main activated deformation components during data-driven deformation. We use the same control points manually chosen by the user (8 vertices on the $29^{\rm th}$ shape in the SCAPE~\cite{anguelov2005scape}) to reconstruct the shape with the data-driven deformation method~\cite{gao2017sparse} and the same number of components. The figure shows that our result is more plausible than the existing methods, and it is similar to the ground truth.}}%
\label{fig:reconbycontrol}
\end{figure}

\begin{figure*}[ht]
\begin{center}
\includegraphics[width=0.9\linewidth]{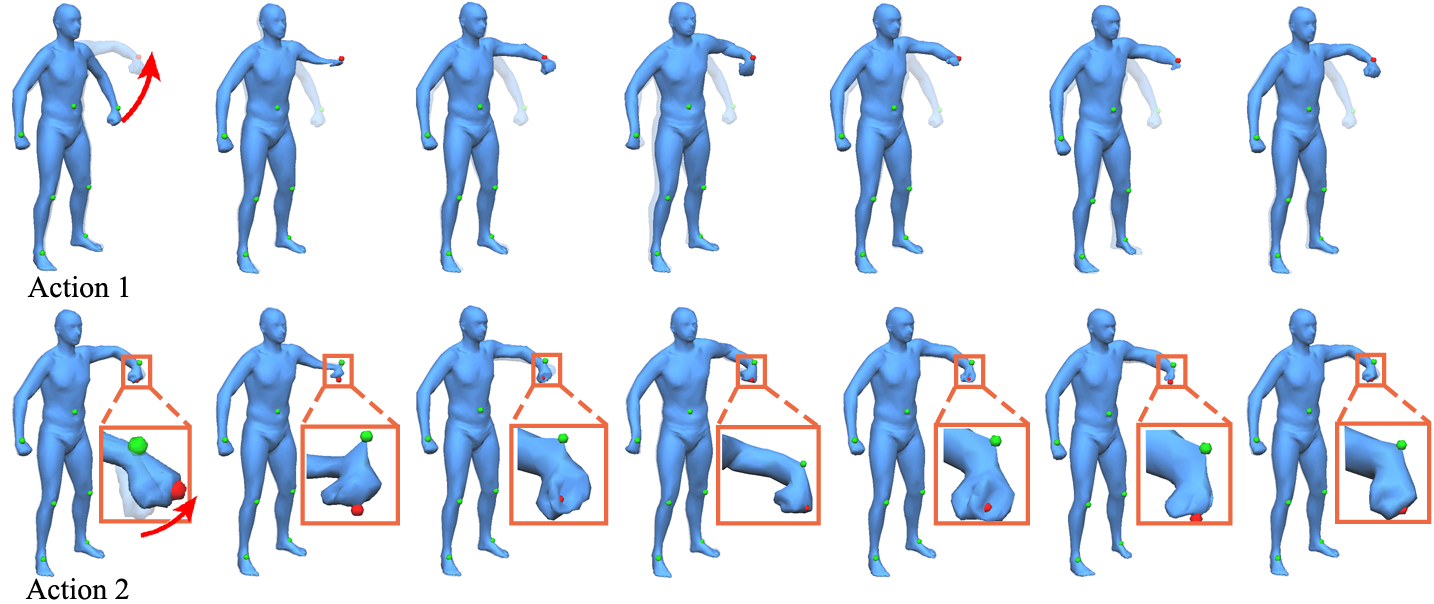}\\
\includegraphics[width=0.9\linewidth]{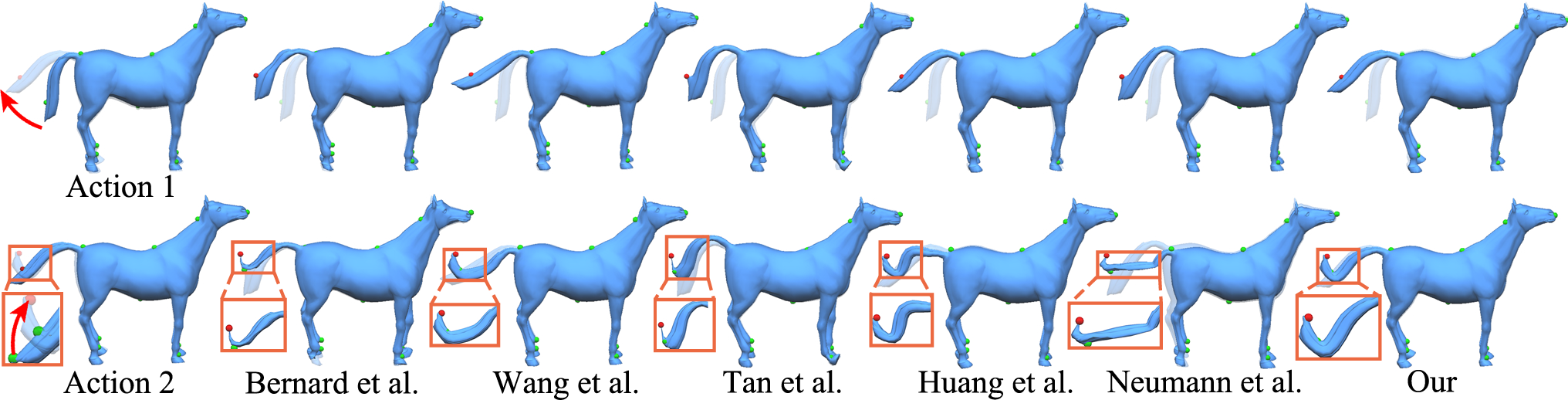}
\end{center}%
\caption{Comparison of multiscale shape editing. We compare the editing results with  different methods on the SCAPE~\cite{anguelov2005scape} and Horse~\cite{Sumner:2004:DTT:1015706.1015736} datasets. The results demonstrate that our extracted deformation components are suitable for multiscale shape editing. The first column shows the editing steps. Every row gives the deformed results of the corresponding action of various methods. The differences with other methods are highlighted in the orange rectangle with closeups to show the details. The existing methods have obvious distortions, demonstrating the superiority of our multiscale deformation components.}%
\label{fig:multiscale}
\end{figure*}

\subsection{Qualitative Evaluation}%

We also provide qualitative evaluation of extracted multiscale deformation components, by comparing the visualization results of our method with the other methods.
\yjrev{In our experiments, we have two level autoencoders. For the each level autoencoders, we extract corresponding scale deformation components from the fully connected layer using the same method as~\cite{sparsevae2017}. For every autoencoders, we can extract same number components as the dimension of latent space. So we visualize each extracted components in each level. Due to our setting ($K_z = [10,5]$), the first level and second level autoencoders output 10 and 5 deformation components respectively. For all results, there are some slight deformation components that is not shown for the brief and reasonable visualization. We remove these components by the post-process (Sec.4.3). Furthermore, we colored deformed region compared with the reference mesh.}

\yjtvcg{As independently extracted deformation components do not correspond to each other, we adopt the visualization method in~\cite{huang2014sparse,sparsevae2017,wang2017articulated}, and manually select two components with similar deformation areas as much as possible.} Figs.~\ref{fig:leftarm} and~\ref{fig:horsecomp} show the comparison with various methods on the SCAPE and Horse datasets. We show the corresponding deformation components in a similar area on the shape (left arm of the SCAPE and every major part of the horse). Our results are more meaningful, capturing major deformation modes in a multiscale manner.

\yj{We then verify if our extracted components are more semantically meaningful than other methods by a user study. We test it on three datasets (SCAPE, Horse and Pants) and compare our method with five state-of-the-art methods~\cite{bernard2016linear,huang2014sparse,wang2017articulated,sparsevae2017,neumann2013sparse}. \yjtvcg{We adopt a scoring method to determine whether a component has a certain semantics meaningfulness. The scoring is based on the participants’ perception of semantic appropriateness of each component.} For each data set, we first render the components extracted by all methods into the same rendering style and mix them together. We let the users browse all the pictures \yjtvcg{to have an idea of distribution} first, and then let the users score in the range from 0 to 100 for each rendered image using a slider. 10 participants were involved in the user study. 
Then we count the average score of each method by every participant as shown in Table~\ref{tab:userstudy}. Our method receives highest scores in all the datasets.

In summary, compared with existing methods, our method can extract plausible and reasonable localized deformation components with semantic meanings, while the other methods have some distortions and cannot extract multiscale deformation components. \yjtvcg{Note that our extracted components correspond to \emph{deformations}, rather than semantic \emph{parts}, so it is natural that they are not always aligned with semantic segmentation, but instead aligned at the motion sequence level, as observed also in previous works~\cite{bernard2016linear,huang2014sparse,neumann2013sparse,sparsevae2017,wang2017articulated}.}
}
\begin{table}[!t]
\fontsize{9}{12}\selectfont
\tabcaption{\yj{The user study that verifies the semantic meaning of deformation components extracted by various methods. We ask 10 participants and report their average scores of each method on the three datasets.}}
\label{tab:userstudy}
\begin{tabular}{p{0.8cm}<{\centering}p{1.1cm}<{\centering}p{0.9cm}<{\centering}p{0.8cm}<{\centering}p{0.8cm}<{\centering}p{0.7cm}<{\centering}p{0.7cm}<{\centering}}
\toprule
\multirow{2}{*}{DataSet}&Neumann  &Bernard &Wang &Huang  & Tan & \multirow{2}{*}{Our}\\
& et al. &  et al. & et al. &et al. &et al.&  \\
\hline
Horse& $46.69$ & $30.37$ & $48.37$ & $56.28$ & $52.41$ & $\mathbf{68.00}$ \\
\hline
SCAPE& $33.87$ & $40.87$ & $55.29$ & $48.13$ & $61.33$ & $\mathbf{72.54}$ \\
\hline
Pants& $36.37$ & $55.75$ & $23.00$ & $50.50$ & $46.00$ & $\mathbf{73.37}$\\
\bottomrule
\end{tabular}

\end{table}

\yjtvcg{In addition, we show more visualization results of multiscale deformation components on the following datasets: Swing~\cite{Vlasic2008}, a fat person (ID: 50002) from Dyna~\cite{Dyna:SIGGRAPH:2015}, Flag, Dress and Skirt. \yjrev{The Flag, Dress and Skirt dataset are synthesized by physical simulation. The Skirt dataset contains more complex motion and deformations. In Figs.~\ref{fig:fattree}, \ref{fig:swingtree}, \ref{fig:dresstree}, \ref{fig:dresstree2}, \ref{fig:flagtree}, the components extracted by our method are in a multiscale manner and consistent with semantic meanings and our method can learn deformation components of different scales with multiscale structure in a complex data}. 
}

We further compare shape editing using various methods by given control points and deformation components they extract. An example is shown in Fig.~\ref{fig:reconbycontrol}. We use the 8 control points (rendered as green balls) on the $29^{\rm th}$
shape in the  SCAPE dataset~\cite{anguelov2005scape} manually chosen by the user. \yj{Then we apply the data-driven method~\cite{gao2017sparse} 
to reconstruct it with the help of the extracted deformation components by various methods. As the \yjtvcg{left part} of Fig.~\ref{fig:reconbycontrol} shows, our result is similar to the ground truth and plausible, while the other results have some artifacts and distortions, such as the right arm of~\cite{wang2017articulated} and left arm of~\cite{sparsevae2017}. The right part of Fig.~\ref{fig:reconbycontrol} shows the three main activated components during data-driven deformation.}
Since the SCAPE contains much large-scale rotation, the method~\cite{sparsevae2017} 
only focuses on extracting large-scale deformation, but fails to capture important fine details, which results in the serious distortion in the arm due to lack of essential components.

\subsection{\yjtvcg{Parameter Settings and Ablation Study}}\label{sec:ablation}
\yj{In this section, we evaluate the model sensitivity to the parameters, including the weights \yjtvcg{($\mathbf{\lambda_1}, \mathbf{\lambda_2}$) in the loss function, the size of deformation region ($d$ of $\Lambda_{ik}$)}, the effect of attention mechanism on generalization error and the difference between joint training and separate training. 

\begin{figure}%
\begin{center}
\includegraphics[width=0.9\linewidth]{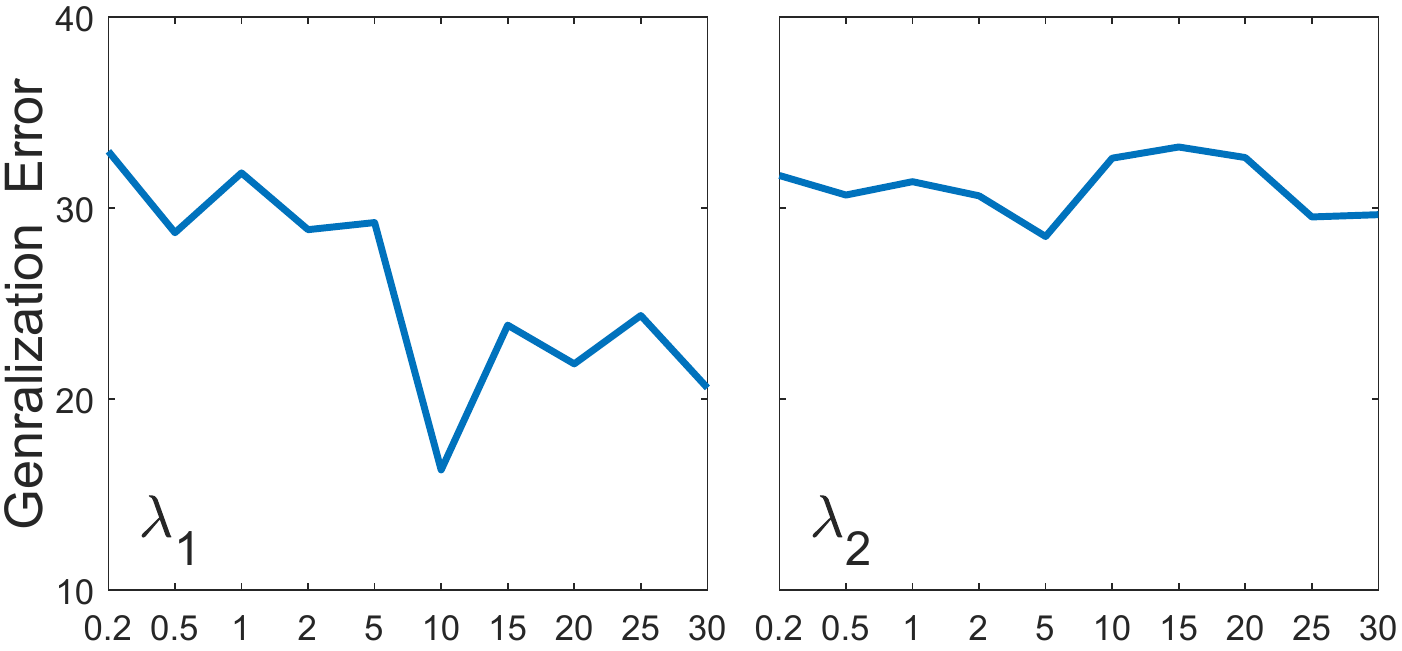} 
\end{center}
\centering
\caption{$E_{rms}$ errors of generating unseen shapes with our overall autoencoder w.r.t. the weights $\lambda_1$ and $\lambda_2$. The figure shows that our network can get lower errors when $\lambda_1=10$ and is robust to different choice of $\lambda_2$.}
\label{fig:weight}
\end{figure}

\begin{figure}%

\centering
\includegraphics[width=0.9\linewidth]{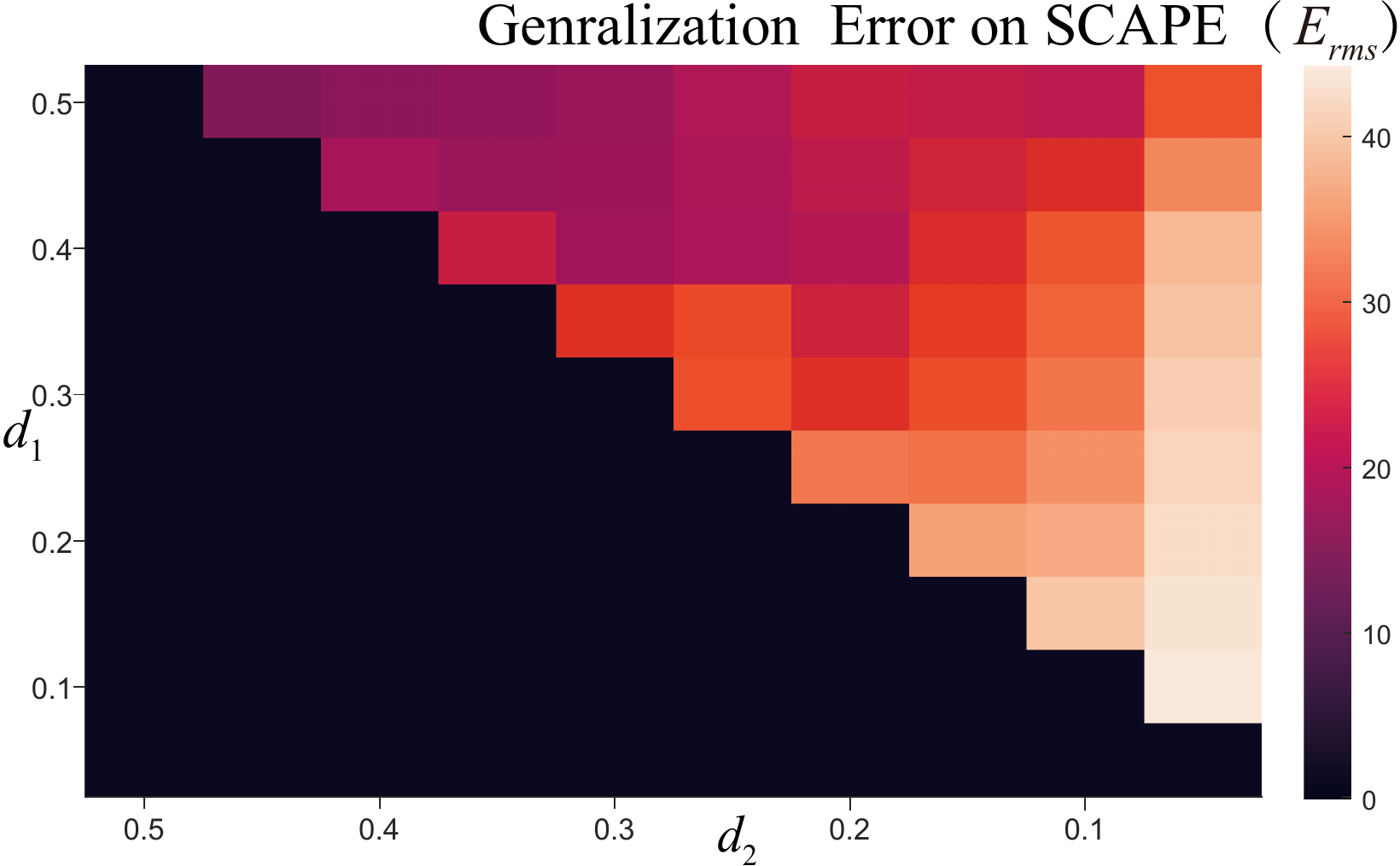} %
\caption{The relationship between $E_{rms}$ of generating the unseen data and ($d_1$, $d_2$) is visualized, where $d_1$ and $d_2$ are from $AE_0$ and $AE_k,1\le k \le K$ on SCAPE data. The noise in the figure is a result of the gradient-descent training procedure. The black (below the main diagonal) means no data, as by definition $d_1$ should be larger than $d_2$.}
\label{fig:d}
\end{figure}

\begin{table*}[]
\fontsize{9}{12}\selectfont
\tabcaption{\yjrev{Comparison of different training strategies and the influence of the attention mechanism for shape reconstruction. We show results comparing joint training with separate training, and whether or not the attention mechanism is used.
For each setting, we test it on several data sets by computing the $E_{rms}$ of the reconstructed shapes for unseen shapes. It shows that joint training with attention mechanism give the best results.}}
\label{tab:eval_train_att}
\begin{center}
\begin{tabular}{p{2.5cm}<{\centering}p{1.1cm}<{\centering}p{1.1cm}<{\centering}p{1.0cm}<{\centering}p{1.2cm}<{\centering}p{1.0cm}<{\centering}p{1.1cm}<{\centering}p{1.1cm}<{\centering}p{1.1cm}<{\centering}p{1.0cm}<{\centering}p{1.0cm}<{\centering}}
\toprule
DataSet & Swing & Scape & Pants & Humanoid & Horse & Flag & Dress & Jump & Fat & Face \\ \hline
\multirow{2}{*}{\tabincell{c}{With Attention\\Joint Training}}& \multirow{2}{*}{$\mathbf{12.2615}$} & \multirow{2}{*}{$\mathbf{23.6807}$} & \multirow{2}{*}{$\mathbf{6.4083}$} & \multirow{2}{*}{$\mathbf{4.1538}$} & \multirow{2}{*}{$\mathbf{6.9246}$} & \multirow{2}{*}{$\mathbf{20.0627}$} & \multirow{2}{*}{$\mathbf{11.5744}$} & \multirow{2}{*}{$\mathbf{16.3475}$} & \multirow{2}{*}{$\mathbf{4.3465}$} & \multirow{2}{*}{$\mathbf{1.4409}$} 
\\ &&&&&&&&&& \\
\hline
\multirow{2}{*}{\tabincell{c}{Without Attention\\Joint Training}}& \multirow{2}{*}{$17.1974$} & \multirow{2}{*}{$46.2339$} & \multirow{2}{*}{$8.2945$} & \multirow{2}{*}{$6.0285$} & \multirow{2}{*}{$20.8894$} & \multirow{2}{*}{$25.2675$} & \multirow{2}{*}{$12.5284$} & \multirow{2}{*}{$23.2829$} & \multirow{2}{*}{$5.3045$} & \multirow{2}{*}{$2.500$} 
\\ &&&&&&&&&& \\
\hline
\multirow{2}{*}{\tabincell{c}{With Attention\\Separate Training}}& \multirow{2}{*}{$20.2918$} & \multirow{2}{*}{$36.1484$} & \multirow{2}{*}{$15.0249$} & \multirow{2}{*}{$6.0469$} & \multirow{2}{*}{$11.5899$} & \multirow{2}{*}{$24.4949$} & \multirow{2}{*}{$14.0248$} & \multirow{2}{*}{$19.1332$} & \multirow{2}{*}{$4.4469$} & \multirow{2}{*}{$2.1345$} 
\\ &&&&&&&&&& \\
\hline
\multirow{2}{*}{\tabincell{c}{Without Attention\\Separate Training}}& \multirow{2}{*}{$26.7981$} & \multirow{2}{*}{$43.4439$} & \multirow{2}{*}{$19.8732$} & \multirow{2}{*}{$8.0583$} & \multirow{2}{*}{$17.4221$} & \multirow{2}{*}{$28.9938$} & \multirow{2}{*}{$17.3392$} & \multirow{2}{*}{$24.9437$} & \multirow{2}{*}{$7.3341$} & \multirow{2}{*}{$3.9472$} 
\\ &&&&&&&&&& \\
\bottomrule
\end{tabular}
\end{center}

\end{table*}

\subsubsection{The choice of $\mathbf{\lambda_1},\mathbf{\lambda_2}$}
We test the influence of parameters $\lambda_1,\lambda_2$ on the generalization ability of the network. We evaluate it by  $E_{rms}$ errors of reconstructing unseen shapes on the SCAPE dataset. By fixing $\lambda_1$ to the default value $10$, we change $\lambda_2$ from $0.2$ to $30$ as the right curve of Fig.~\ref{fig:weight} shows. The result shows our network is robust to different choice of $\lambda_2$. With fixed $\lambda_2$, we change $\lambda_1$ from $0.2$ to $30$ as the left curve of Fig.~\ref{fig:weight} shows. The result justifies that our network can get lower errors when $\lambda_1=10$, which is chosen as the default value in our experiments.

\subsubsection{The choice of $d$ in $\mathbf{\Lambda_{ik}}$} 
The other parameter to choose is $d$ in $\Lambda_{ik}$, which is the parameter that determines the scale of the extracted components. \yjrev{$d_1, d_2$ are cutoffs for normalized geodesic distance with range from 0 to 1, thus, the original size of specific dataset has little affect on them. These parameters only reflect the related size of the localized deformation components compared to the whole model.

In our network, by default we stack two levels of autoencoders, so we need choose two parameters $d_1,d_2$ for both the first-level and the second-level autoencoders ($AE_0$ and $AE_1$) respectively. In order to extract multiscale deformation components, we need ensure $d_1>d_2$, then $AE_1$ will cover \hyh{a more detailed} region than $AE_0$. 
As Fig.~\ref{fig:d} shows, we test the network generalization ability ($E_{rms}$) on the SCAPE dataset with different combinations $d_1,d_2$ by changing values from $0.05$ to $0.5$ with step $0.05$. The figure illustrates that smaller $d_1$ and $d_2$ result in larger reconstruction errors.
\hyh{Although lower error can be achieved when $d_1$ and $d_2$ are large enough (close to 0.5), we will extract more global deformation components (i.e., not localized)}
It is a trade off progress and we must balance the generalization ability and the extracted deformation components. In our experiments, we set $d=[0.4,0.2]$ that can make network extract localized components and keep lower reconstruction error simultaneously. }

\subsubsection{The number of level of AE}\label{sec:ae_levels}
\gl{The architecture of this neural net supports multiple levels of deformation scales from coarse to fine. In almost all tested dataset in this paper, the two level AEs architecture is enough to represent. }
The error between the input ACAP feature $X_i$ and the reconstructed ACAP feature $\hat{X_i}, 1<i<N$ is divided by the norm of $X_i$ to get the percentage as shown in Eqn.~\ref{eqn:percentage}. 

\begin{equation}\label{eqn:percentage}
Perc_{i} = \frac{||X_i-\hat{X_i}||_F^2}{||X_i||_F^2}
\end{equation}
\gl{where the $||\cdot||_F$ is the Frobenius norm, $Perc_{i}$ is the percentage of error on the $i^{\rm th}$ shape in this dataset. Then, we choose the maximum percentage of error in the whole dataset as the proportion of deformation that has not been represented. We show the results in the Table~\ref{tab:ae_levels}. The experiment illustrates that the percentage of two-level AEs is very low in all datasets so that our two-level AEs is enough.}

\begin{table*}[]
\fontsize{9}{12}\selectfont
\tabcaption{\gl{The maximum percentage of the reconstruction error for each shape in every tested dataset in this paper. The values are calculated by the Eqn.\protect\ref{eqn:percentage}, as shown in table, the two-level AEs could represent the deformation of the  very well with the tiny percentage of the reconstruct error, while the reconstruction error with one-level AE is much larger for reconstructing the input feature. 
}}
\label{tab:ae_levels}
\begin{center}
\begin{tabular}{cccccccccc}
\toprule
DataSet & Horse & Face & Jumping & Swing & Pants & Dress & Fat & Flag \\ \hline
Percentage of Error (one-level AE) & 0.1219 & 0.06325 & 0.2312 & 0.4756 & 0.1763 & 0.5471 & 0.1322 & 0.1983\\ \hline
Percentage of Error (two-level AEs) & 0.005491 & 0.001065 & 0.007800 & 0.06876 & 0.01745 & 0.06296 & 0.005652 & 0.02674 \\
\bottomrule
\end{tabular}
\end{center}
\end{table*}

\yjrev{
\subsubsection{The choice of $K_z$}
The $K_z$ describes the dimension of latent space of two level AEs. The hyper-parameter decides the number of deformation components in each autoencoder of each level. Since methods compared in the paper produce 50 deformation components, we let $K_{z_0} \times K_{z_1} = 50$. We have the following combinations: $K_z = [1, 50], [2, 25], [5, 10], [10, 5], [25, 2], [50, 1]$. However, the $[1, 50], [50, 1]$ are the trivial setting. So we test the other four settings of $K_z$ on the SCAPE dataset. The Table~\ref{tab:K} shows the results, which illustrate that it perform well when $K_z = [25, 2]$. But, there are only two deformation components for each second level AEs, which is not reasonable for most datasets. For example, in Fig.~\ref{fig:scapetree}, our network extract three meaningful components for some brunches on SCAPE dataset. Furthermore, most of our experimental results show that there are more than 2 meaningful components extracted from the second level AEs. So we choose the second higher performance with the setting $k_z = [10, 5]$.
 
\begin{table}[]
\centering
\tabcaption{\gl{The reconstruction errors on the unseen data from SCAPE with different settings of $K_z$. For each setting, it represents the dimension of latent space of two level AEs. From the table, it shows that our method has high performance on the setting $K_z=[25, 2]$. However, there are only up 2 deformation components for each second level AEs with the setting, which is not reasonable for most of datasets. For example, in figure~\ref{fig:scapetree}, our network extract three meaningful components for some brunches on Scape dataset. Furthermore, most of our experimental results show that there are more than 2 meaningful components extracted from the second level AEs. So we choose $K_z = [10, 5]$.
}}
\label{tab:K}
\begin{tabular}{ccccc}
\toprule[1pt]
$K_z$ & $[2, 25]$ & $[5, 10]$ & $[10, 5]$ & $[25, 2]$  \\ \hline\hline
$E_{rms}$ & 34.9763 & 29.1127 & 23.6807 & 21.7749\\
\bottomrule[1pt]
\end{tabular}
\end{table}

}

\subsubsection{Training strategies and attention mechanism}
Finally, we evaluate the effect of the attention mechanism and different training strategies. The statistics are shown in  Table~\ref{tab:eval_train_att} based on the experiments on the SCAPE dataset. 

For the training strategy, we compare the reconstruction error ($E_{rms}$) by training the network either jointly or separately. The results are shown in the first and third rows of Table~\ref{tab:eval_train_att}, and jointly training the network can get better results. %

\yjrev{We also perform the experiment to demonstrate the effect of the attention mechanism qualitatively and quantitatively. 
The results are shown in the first and second rows of  Table~\ref{tab:eval_train_att}. Training with attention mechanism will get lower errors. The reason is that each autoencoder of the second-level will focus on a different sub-region to minimize the loss.

For the qualitative evaluation, we show the results in Fig.~\ref{fig:dresstree}. If we train our network without the attention mechanism, our network architecture degenerates into the version of Tan et al.~\cite{sparsevae2017} with two level \hyh{autoencoders}. In the second level AEs, there will be only \hyh{one} AE because all of them are same. \hyh{Despite this, different level autoencoders are also able to extract the different scale deformation components as shown in the right part of Fig.~\ref{fig:dresstree}}. In the left part of Fig.~\ref{fig:dresstree}, the second and third row represent the deformation components that are extracted by first level AE and second level AE respectively, which is processed by removal of redundant components. The first row is the some sample data in Dress dataset. The results illustrate that the deformation components no longer have a multiscale structure \hyh{when the second level AEs do not focus on sub-regions to extract the deformations components.}}
}

\subsection{Multiscale Shape Editing}%
Multiscale shape editing is an important application in computer graphics. Users usually start with editing of  the overall shape, and then focus on adjusting  the details. With existing methods, the extracted deformation components either contain some global information~\cite{neumann2013sparse,bernard2016linear} thus making the components unsuitable for local editing,
\yjrev{or focusing too much on large-scale deformations and failing to capture essential small-scale deformations for faithful reconstruction, leading to distortions like~\cite{sparsevae2017}, }
which would affect the users' editing efficiency for 3D animations. \yjtvcg{Given a shape deformation dataset that contains %
diverse deformations, our method can produce multiscale localized  deformation components which are visually \yjtvcg{semantically} meaningful,
corresponding to typical deformation behaviour. Along with data-driven deformation~\cite{gao2017sparse}, this allows users to edit shapes efficiently and intuitively under the constraints of the control points and subspace spanned by extracted deformation components. Please refer to the work~\cite{gao2017sparse} for details of data-driven deformation.} 

Fig.~\ref{fig:multiscale} shows some examples.
\yj{For the SCAPE dataset, we design two actions: raising the left arm (Action 1) and then turning the wrist (Action 2). All the compared methods perform well in Action 1. However, in Action 2, only our method can naturally twist wrist with the help of our extracted multi-scale deformation components. In contrast, all other methods have various distortions. For the Horse dataset, we also design two actions: raising the whole tail (Action 1) and then twisting the end of the tail (Action 2). Our method can bend the end of the tail naturally after raising the whole tail. But other methods have more distortions and even lead to changes on the entire tail, especially the methods~\cite{neumann2013sparse,bernard2016linear} based on the Euclidean coordinate representation. In summary, our extracted multiscale deformation components can perform better than existing methods in multiscale shape editing.} \error{See the accompanying video for more results.}

\section{Limitation and Conclusion}\label{sec:conclusion}
In this paper, we propose a novel autoencoder with the attention mechanism to extract multiscale localized deformation components. \yjtvcg{We use stacked AEs to extract multiscale deformation components. This helps capture richer information for better shape editing, with better generalization ability (see Table~\ref{moreevaluation}). Moreover, the first-level AE extracts some coarse level components and learns attention masks to help the second-level AEs focus on relevant sub-regions to extract fine-level components.} Extensive quantitative and qualitative evaluations show that our method is effective, outperforming state-of-the-art methods. The extracted deformation components by our method can be used on multiscale shape editing for computer animation, \yjtvcg{which demonstrates that the extracted multiscale localized deformation components are effective and meaningful for reducing the user's efforts. In the future, this work can give more solutions and explorations on applying the attention mechanism on 3D shape analysis and synthesis.

\yjrev{Although our method can analyze the shape dataset with multiscale manner to extract some different-scale components for easy shape editing, there are some limitations. While datasets including ShapeNet contains the general 3D shapes, our methods cannot handle those kinds of data, especially for CAD manufacturing. The data that our method can handle must be with the same connectivity. \hyh{Also, our method uses} a fixed network architecture and attention mechanism to analyze the shape with multiscale manner, so the multiscale structure of shape set will be similar and obey the fixed network structure for all dataset. We cannot capture the multiscale variation between different categories in the dataset. \hyh{In the future, We are committed to} exploit a method that can analyze and capture the variation of the multiscale structure on ShapeNet automatically. Finally, we can also merge the post-process (Sec~\ref{sec:removal}) to the neural network which can predict the number of sub-parts automatically.}}

\section*{Acknowledgment}

\yjtvcg{
This work was supported by National Natural Science Foundation of China~(No. 61872440 and No. 61828204), Beijing Municipal Natural Science Foundation~(No. L182016),  Royal Society Newton Advanced Fellowship  (No. NAF$\backslash$R2$\backslash$192151), Youth Innovation Promotion Association CAS, CCF-Tencent Open Fund and Open Project Program of the National Laboratory of Pattern Recognition (No. 201900055).
}

\ifCLASSOPTIONcaptionsoff
  \newpage
\fi

\bibliographystyle{IEEEtran}
\bibliography{IEEEabrv,./egbib}

\end{document}